\documentclass[usenatbib]{mn2e}
\usepackage{graphicx}
\usepackage{amssymb}
\usepackage{longtable}
\usepackage{dpfloat}
\usepackage{subfig}
\usepackage{journal_names}

\title[Global Properties of `Ordinary' ETGs]{Global Properties of `Ordinary' Early-type Galaxies: photometry and spectroscopy of stars and globular clusters in NGC~4494}

\author[C. Foster et al.]{Caroline Foster,$^1$\thanks{E-mail: cfoster@astro.swin.edu.au} Lee R. Spitler,$^1$ Aaron J. Romanowsky,$^2$ Duncan A. Forbes,$^1$ 
\newauthor Vincenzo Pota,$^1$ Kenji Bekki,$^3$ Jay Strader,$^{4}$ Robert N. Proctor,$^5$ Jacob A. Arnold,$^2$ 
\newauthor and Jean P. Brodie$^2$
\\
$^1$Centre for Astrophysics \& Supercomputing, Swinburne University, Hawthorn, VIC 3122, Australia\\
$^2$UCO/Lick Observatory, University of California, Santa Cruz, CA 95064, USA\\
$^3$ School of Physics, University of New South Wales, Sydney, NSW, 2052, Australia\\
$^4$Harvard-Smithsonian Center for Astrophysics, 60 Garden St., Cambridge, MA 02138, USA\\
$^5$Universidade de S\~ao Paulo, IAG, Rua do Mato 1226, S\~ao Paulo 05508-900, Brazil}

\begin{document}

\maketitle

\begin{abstract}
We present a comprehensive analysis of the spatial, kinematic, and chemical properties of stars and globular clusters (GCs) in the  `ordinary' elliptical galaxy NGC~4494 using data from the Keck and Subaru telescopes. We derive galaxy surface brightness and colour profiles out to large galactocentric radii. We compare the latter to metallicities derived using the near-infrared Calcium Triplet. We obtain stellar kinematics out to $\sim3.5$ effective radii. The latter appear flattened or elongated beyond $\sim 1.8$ effective radii in contrast to the relatively round photometric isophotes. In fact, NGC~4494 may be a flattened galaxy, possibly even an S0, seen at an inclination of $\sim45$ degrees. We publish a catalogue of 431 GC candidates brighter than $i_0=24$ based on the photometry, of which 109 are confirmed spectroscopically and 54 have measured spectroscopic metallicities. We also report the discovery of 3 spectroscopically confirmed ultra-compact dwarfs around NGC~4494 with measured metallicities of $-0.4\lesssim[Fe/H]\lesssim-0.3$. Based on their properties, we conclude that they are simply bright GCs. The metal-poor globular clusters are found to be rotating with similar amplitude as the galaxy stars, while the metal-rich globular clusters show marginal rotation. We supplement our analysis with available literature data and results. Using model predictions of galaxy formation, and a suite of merger simulations, we find that many of the observational properties of NGC~4494 may be explained by formation in a relatively recent gas-rich major merger. Complete studies of individual galaxies incorporating a range of observational avenues and methods such as the one presented here will be an invaluable tool for constraining the fine details of galaxy formation models, especially at large galactocentric radii.
\end{abstract}


\section{Introduction}\label{sec:intro}
\defcitealias{V03}{V03}

Despite sustained efforts both on the observational and theoretical fronts, an accepted self-consistent picture of galaxy formation and evolution has not yet emerged. Key processes involved during galaxy formation and evolution include (but are not limited to) galaxy mergers, the accretion and/or dissipation of gas (whether early, via monolithic collapse, or merger induced), feedback processes such as stellar winds, supernova (SN) feedback or active galactic nuclei (AGN), reionisation, etc.

Observational clues to the importance of these processes are crucial to our understanding of galaxy formation. For example, the relative importance of gas dissipation and energy feedback from stellar winds, SN and AGN as star formation quenching mechanisms needs to be constrained observationally. 
%Gas dissipation gives rise to central star formation, and hence metal enrichment. On the other hand, if gas is ejected via some feedback mechanism(s), star formation is inhibited, and with it, metal enrichment. Stellar winds and SN feedback eject gas more efficiently at large galactocentric radii \citep{Matteucci94,Martinelli98}, while AGN are most efficient at ejecting gas from the centre and sometimes out to large radii \citep{Croton06,Wang08}. Therefore, 
In principle, this can be studied through looking at radial abundance gradients in a galaxy \citep[e.g.,][]{Bekki99,Kobayashi99,Hopkins09}.

%Also, the relative roles played by the mass ratio of the progenitors involved in a galaxy merger (i.e., minor- versus major-mergers) and other properties of these progenitors (e.g., gas-richness, presence of rotation, etc) in producing the properties of the remnant galaxies observed today need to be understood \citep[e.g.,][]{LopezSanjuan10,Hopkins09,Hopkins10}. 
Also, the properties of progenitor galaxies involved in a merger can be probed by looking for distinct kinematic and morphological signatures in the merger remnant galaxy \citep[e.g.,][]{LopezSanjuan10,Hopkins09,Hopkins10}, especially at large galactocentric radii. Indeed, \citet{Hoffman10} predict that the less relaxed intermediate and outer parts of a gas-rich (or wet) disk-disk merger remnant may retain the dynamical signature of the cold disk stars and original halo stars, respectively. 
%The stellar properties within the inner $\lesssim1$ effective radius ($r_e$) are shaped by stars recently formed through gas dissipation. 
In this model, a transition in the kinematic properties of the remnant is expected for $1\lesssim r_e\lesssim3$ where $r_e$ is the effective radius. % beyond which, the kinematics resembles that of a dissipationless (or dry) merger. 
%This is because the outer parts should be built up from the progenitor's pre-existing disk and halo stars in both cases. 
Moreover, as a complement to looking for these kinematic signatures, dry merging also leaves an imprint on the stellar populations of the remnant such that it washes-out or weakens metallicity gradients \citep[e.g.,][]{White80,DiMatteo09,Pipino10}.

Thus, in order to obtain a complete picture and understanding of galaxy formation it is necessary to spectroscopically probe both the global kinematics and stellar populations present in galaxies out to large galactocentric radii  (i.e., $>1r_e$). Unfortunately, the low surface brightness of galaxies in the outskirts hinders spectroscopic studies at large galactocentric radii. Spatially resolved kinematic and stellar population studies of the stellar light of large early-type galaxies (ETGs) are typically confined to the inner $\lesssim 1r_e$ \citep[e.g.,][]{Reda07,Emsellem07,Emsellem11,Kuntschner10}, thereby probing less than half the stellar light and potentially missing the important transition region previously discussed. To remedy this, \citet[][hereafter P09]{Proctor09}\defcitealias{Proctor09}{P09} have developed a technique that takes advantage of the large field-of-view of the DEIMOS spectrograph on Keck to extract galaxy light spectra from background spectra (sky $+$ galaxy light) of GC multi-slit observation and derive spatially resolved kinematics out to 3$r_e$ \citepalias{Proctor09} and metallicities out to $\sim1.5r_e$ \citet[][hereafter F09]{Foster09}.\defcitealias{Foster09}{F09}

%Without the knowledge of the precise star formation history of a given galaxy, probing the stellar populations of the unresolved stellar light yields only the luminosity weighted average population. Because younger stars tend to outshine old ones, the inferred stellar population parameters are disproportionately biased towards those of the most recent star formation episode and may not represent the bulk of the stars \citep{Conroy09,Conroy10}. However, ETGs tend to be dominated by old stellar populations ($\sim5$ Gyr and older). Nevertheless, this effect must be kept in mind as a caveat of any stellar population study of unresolved galaxy light.

A good probe of early star formation are globular clusters (GCs). ETGs typically present a larger specific frequency (i.e., number per unit mass) of globular clusters (GCs) than late-type galaxies \citep[e.g.,][]{vandenBergh82,Ashman92}. GCs are generally measured to be old with ages comparable to the age of the Universe \citep[e.g.,][]{Brodie05,Strader05,Puzia05,Cenarro07,Norris08,Proctor08} and thus probe the very earliest star formation episodes of their host galaxy. This known old age significantly diminishes the difficulties associated with the age-metallicity degeneracy \citep{Worthey94a} wherein old metal-poor populations share certain photometric and spectroscopic properties with those of young metal-rich populations. Moreover, GCs are well approximated as single-stellar populations, which greatly simplifies their stellar population analysis and interpretation.

Understanding GC formation is crucial to understanding galaxy formation, especially the early epochs \citep{West04}. For example, the nearly ubiquitous colour bimodality of the GC systems of most galaxies \citep[see][for a review]{Brodie06} is usually interpreted as a metallicity bimodality with the blue and red GCs being metal-poor and -rich, respectively. This is a stringent constraint for galaxy formation scenarios that must provide at least two formation episodes or mechanisms to explain the colour bimodality (e.g., \citealt{Lee10}; but see \citealt{Muratov10}). 

A recent and extensive GC study is that of the kinematics and stellar populations of a statistically sizeable fraction of the GC system of the nearby giant ETG NGC~5128 by \citet[][a,b]{Woodley10b}. They find that the majority of the GCs in both subpopulations are old with a significant population of young metal-rich GCs forming later. They conclude that these young metal-rich GCs may have been formed in a more recent merging event than the bulk of the GCs. Moreover, both metal-rich and -poor GC subpopulations are found to be pressure supported with only mild rotation for the metal-rich GCs. From this, they are able to infer that the GC system of NGC~5128 is consistent with a hierarchical formation in a scenario similar to that proposed by \citet{Beasley02} and \citet{Strader05}. 
%In this scenario, the blue GCs form in an early collapse of small proto-galaxies and are thus expected to be uniformly old as observed. On the other hand, the bulk of the field stars and metal-rich clusters form slightly later during a second collapse following the early hierarchical assembly of the proto-galaxies. The younger metal-rich GCs suggest subsequent major accretion and/or a recent star forming event. This scenario naturally explains the somewhat unorganised motions of both subpopulations. 
This study demonstrates the power of combined spectroscopic and photometric studies of large samples of GCs to constrain the formation and assembly history of individual galaxies.

While the most massive nearby elliptical galaxies are in principle easier to study, their formation may have been atypical due to their often special location at the centre of large potential wells such as galaxy groups or clusters. Therefore, in order to obtain a complete view of galaxy formation, it is necessary to avoid ``special'' environments and aim for more typical galaxy masses (i.e., $\sim M^*$). 
To this end, we focus on the galaxy NGC~4494, which is often described as an ``ordinary elliptical'' galaxy \citep[e.g.,][]{Capaccioli92,Lackner10} mainly based on its typical light profile. Another aspect that makes it fairly ordinary, or average, include its intermediate density environment. Indeed, it has been described in the literature as either isolated \citep{Lackner10} or loose group member \citep{Forbes96,Larsen01} as it is located at the edge of the Coma I cloud. It is neither a large nor a small galaxy with a stellar mass of $\sim10^{11}$ M$_\odot$ (see Table \ref{table:props}). Its elliptical morphology is also typical as early-type galaxies (ETGs) may contain over 50 per cent of the total stellar mass in the local Universe \citep{Bell03}. It contains an inner dust ring \citep[$r<4$ arcsec,][]{Lauer05}, is quite round with an axis ratio of $q=0.87$ (see Table \ref{table:props}) and shows a very smooth luminosity profile with a central cusp \citep{Lauer07}. Peculiarities include a kinematically decoupled core \citep{Bender94} or double maxima \citep{Krajnovic11} in the inner $\sim19$ arcsec beyond which sustained rotation ($V_{\rm rot}\sim 60$ km s$^{-1}$) is observed out to $\sim 3r_e$ with a possible ``pinching'' or ``flattening'' of the kinematics starting at $\sim1.5r_e$ \citepalias{Proctor09}. Other notable properties are a two order of magnitudes deficiency in X-ray luminosity for its optical luminosity \citep{OSullivan04} and a possible deficiency in dark matter \citep[][hereafter N09]{Romanowsky03,Napolitano09}.\defcitealias{Napolitano09}{N09}

Our approach is to study NGC~4494 in great detail to constrain the formation of ``ordinary'' elliptical galaxies. To this end we study its structure, kinematics and stellar populations, as well as its GCs, using imaging and spectroscopy. We also include literature data. Our study is unique and one of the most complete studies of an individual ``ordinary'' elliptical galaxy and its GC system to date. We present arguably the first large catalogue of GC recession velocities in an \emph{ordinary} ETG. 

%Apart from the test of galaxy formation described above, several other scientific problems are tackled with this extensive dataset. For example, the assumption that field stars, planetary nebulae and metal-rich GCs are all tracing the same population is tested by cross-checking their properties for consistency. Moreover, the well-known GC dichotomy usually probed in colour (or metallicity) space is probed in kinematic space.

The paper is divided as follows: Section \ref{sec:data} presents the photometric and spectroscopic data. The surface brightness/density profile, kinematics, colours and metallicity distribution of the stars and GCs are found in Section \ref{sec:analysis}. These results and their implication for the key science questions described above are discussed in Section \ref{sec:discussion}. Finally, we give a brief summary and our conclusions in Section \ref{sec:conclusion}.

\begin{table*}
\begin{tabular}{c c c c c c c c c c c}
\hline\hline
Galaxy & Hubble & $PA_{\rm phot}$& $q_{\rm phot}$  & Distance & $r_{e}$ & $M_{B}$ & $M_{K}$ & Stellar mass & $V_{\rm sys}$ & $\sigma_{0}$\\
 & Type & (deg) &(K band) & (Mpc) & (arcsec) & (mag) & (mag) & ($10^{11}M_{\odot}$) & (km s$^{-1}$) & (km s$^{-1}$)\\
 (1) & (2) & (3) & (4) & (5) & (6) & (7) & (8) & (9) & (10) & (11)\\
\hline
M49& E2 & 163 & 0.81 & $15.1\pm0.7$ & 104 & $-21.5\pm0.16$ & $-25.5\pm0.1$ & $3.1\pm0.3$ & $997\pm7$ & $294\pm3$\\
M60& E2 & 108 & 0.81 & $16\pm1$ & 69 & $-21.2\pm0.2$ & $-25.2\pm0.2$ & $2.4\pm0.4$ & $1117\pm6$ & $335\pm4$\\
M87 & cD & 152 & 0.86 & $15\pm1$ & 95 & $-21.3\pm0.1$ & $-25.1\pm0.1$ & $2.2\pm0.2$ & $1307\pm7$ & $335\pm5$\\
NGC 1399 & E1 & 150 & 1.00 & $19\pm1$ & 80 & $-20.8\pm0.3$ & $-25.0\pm0.2$ & $2.0\pm0.4$ & $1425\pm4$ & $342\pm6$\\
NGC 1407 & E0 & 60& 0.95 & $27\pm3$ & 70 & $-21.4\pm0.3$ & $-25.4\pm0.3$ & $3.0\pm0.9$ & $1779\pm9$ & $272\pm6$\\
NGC 3379 & E1 & 68 & 0.85 & $9.8\pm0.5$ & 35 & $-19.7\pm0.1$ & $-23.7\pm0.1$ & $0.59\pm0.06$ & $911\pm2$ & $209\pm2$\\
NGC 4494 & E1-2 & 173 & 0.87 & $15.8\pm0.9$ & 49 & $-20.4\pm0.2$ & $-24.2\pm0.2$ & $1.0\pm0.2$ & $1344\pm11$ & $150\pm4$\\
NGC 4636 & E0-1 & 143 & 0.84 & $13.6\pm0.9$ & 89 & $-20.2\pm0.2$ & $-24.2\pm0.1$ & $0.95\pm0.08$ & $938\pm4$ & $203\pm4$\\
NGC 5128 & S0 & 43 & 0.89 & $3.9\pm0.3$ & 305 & $-20.2\pm0.2$ & $-24.0\pm0.1$ & $0.78\pm0.07$ & $547\pm5$ & $120\pm7$\\
\hline
\end{tabular}
\caption{Properties of nine massive ETGs with significant GC kinematic samples from the literature. Hubble types (column 2) are as per NASA/IPAC Extragalactic Database (NED).
Position angles and axis ratios (columns 3, 4) are from 2MASS.
Distances (column 5) are based on surface brightness fluctuations \citep{Tonry01} and include the distance moduli correction of \citet{Jensen03}.
Effective radii (column 6) are taken from the Third Reference Catalogue of Bright Galaxies \citep[RC3,][]{deVaucouleurs91} for all galaxies except NGC~5128, which is from \citet{Dufour79}.
$B$- and $K$-band absolute magnitudes (columns 7, 8) are calculated from RC3 and 2MASS apparent magnitudes, respectively, and using the distances quoted in column 5.
Stellar masses (column 9) are calculated from the $K$-band magnitude of column 8 assuming a $M/L_K$ ratio corresponding to the \citetalias{V03} SSP of age 10 Gyrs and solar metallicity.
Systemic velocities (column 10) are from NED.
Central velocity dispersions (column 11) are as per \citet{Paturel03}.}
\label{table:props}
\end{table*}

\begin{figure}
\begin{center}
\includegraphics[width=84mm]{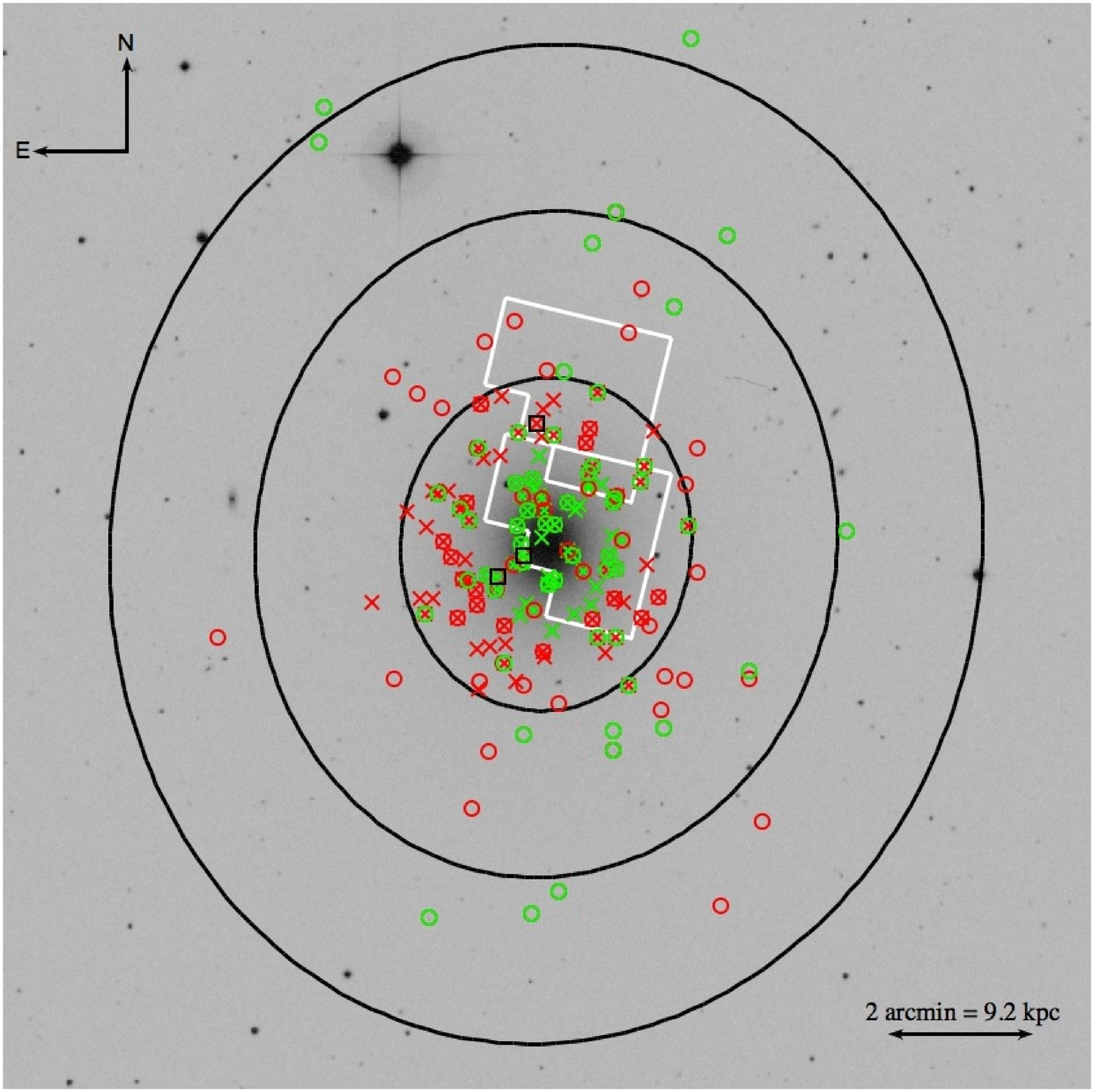}
\caption{DSS image of NGC~4494 showing the positions of our science spectra. Circles and squares represent kinematically confirmed GCs and UCDs, respectively, while crosses show the position of galaxy light spectra. Green and black small symbols are used when the spectrum returned both CaT and kinematic information, while red symbols are used if only kinematic information could be extracted. Large black ellipses represent 3, 6 and 9 $r_e$ with the galaxy's global photometric position angle and axis ratio. The HST field-of-view is shown in white. A colour version of this figure is available in the online version of this article.}\label{fig:slits}
\end{center}
\end{figure}

\section{Data}\label{sec:data}
	\subsection{Imaging acquisition and reduction}\label{sec:phot} 

Subaru Suprime-Cam \citep{Miyazaki02} imaging of NGC~4494 is analysed to understand the GC system and light profiles of NGC~4494. We obtained a $g$-band observation of NGC~4494 during a Gemini time exchange program (GN-2008A-C-12) on the night of 2008 April 2. Two additional bands ($r$ and $i$-bands) were acquired on a later date, 2010 April 4, through Keck time exchange. The total exposure times are 805, 365, and 540 seconds in the $g$, $r$ and $i$-bands, respectively. The seeing conditions in the respective bands are: 0.63, 0.56, and 0.58 arcsecs. Suprime-Cam data are prepared for analysis using standard imaging reduction techniques and a modified version of the {\sc sdfred} data pipeline \citep{Yagi02,Ouchi04}.  The Suprime-Cam field of view covers a $\sim36\times29$ arcmin region centered on NGC~4494.

Suprime-Cam photometry is bootstrapped to the Sloan Digital Sky Survey DR7 \citep{Abazajian09} photometric system using point sources with $19\le i \le21.5$ magnitudes.  The estimated $g$-, $r$- and $i$-band systematic uncertainty due to this calibration is 0.004, 0.005 and 0.004 mags, respectively.  All photometry is Galactic extinction-corrected according to \citet{Schlegel98}.

The images are prepared for GC analysis by first modelling with {\tt IRAF/Ellipse} and subtracting the galaxy light profiles.  {\tt Ellipse} is set to allow the center, position angle and ellipticity to vary. A bright blue star is $\sim6$ arcmin away from NGC~4494 to the NNE. The scattered light from this star extends to a radius of $\sim3.5$ arcmin, thus {\tt IRAF/Ellipse} is also used to model and subtract its light in the $g$, $r$ and $i$-bands.  Because the star and galaxy profiles overlap significantly, {\tt Ellipse} is performed and subtracted iteratively on the light profiles 3 times. The final image products show a constant background value across the field.

A catalogue of GC candidates is constructed from the 3 Suprime-Cam mosaics.  At the distance of NGC~4494, GCs are unresolved and appear as point-sources on the images.  {\tt IRAF/DAOPHOT/FIND} is used to locate objects on the field deviating by $3.8\sigma$ from a global background level.  The difference between two different aperture magnitudes is used as a way to identify and remove extended sources from the GC catalogue \citep[see e.g.][]{Spitler08b}. This is done for detected objects in each image. Aperture corrections are applied to the point sources and the photometric zeropoints derived from the above procedure are used. The three separate photometry catalogues are combined using a matching threshold of 1 arcsec.

To supplement the Suprime-Cam catalogue, an existing Hubble Space Telescope (HST) Wide-field Planetary Camera 2 (WFPC2) GC catalogue is incorporated into the GC analysis.  The WFPC2 catalogue is provided by S. Larsen and is described in \citet{Larsen01}. See Section \ref{sec:selection} for details on our selection of GC candidates.
		
	\subsection{Spectroscopy acquisition and reduction}

\begin{figure}
\begin{center}
\includegraphics[width=84mm]{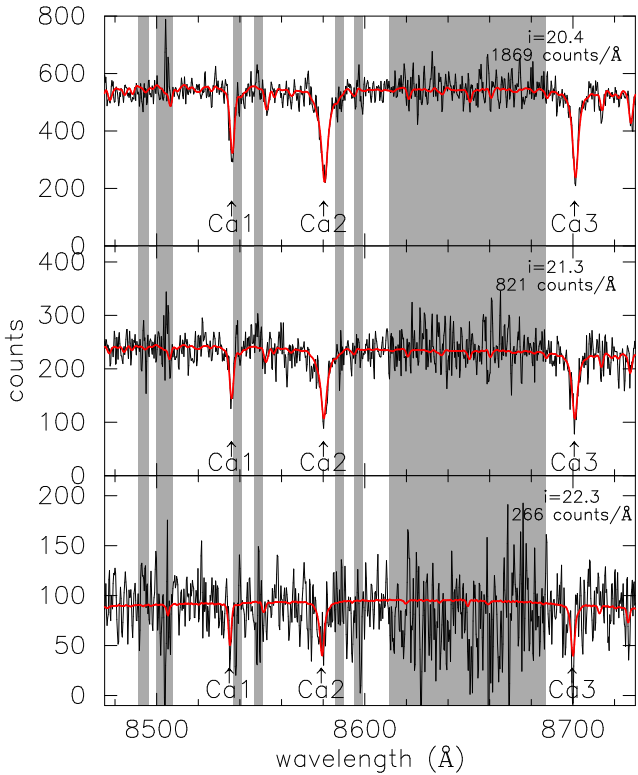}
\caption{Example GC spectra (black) and fitted {\sc pPXF} templates (red) for a bright (top panel), typical (middle panel) and faint (lower panel) GC in our sample. Shaded wavelength bands are regions affected by significant skyline residuals. GC apparent $i_0$ magnitudes are given.}\label{fig:GCspec}
\end{center}
\end{figure}

\begin{figure}
\begin{center}
\includegraphics[width=84mm]{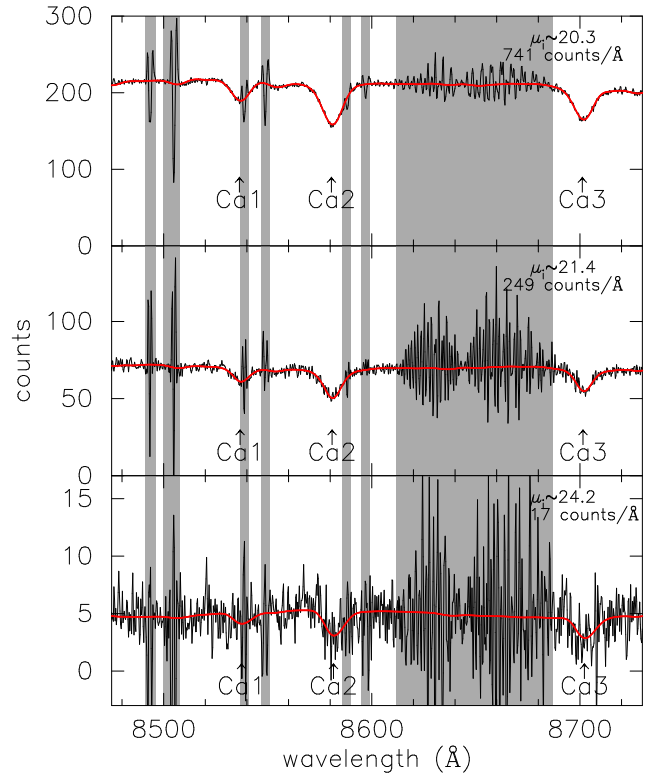}
\caption{Example galaxy light spectra (black) and fitted {\sc pPXF} templates (red) for a high (top panel), typical (middle panel) and low (lower panel) signal-to-noise spectrum. Shaded wavelength bands are regions affected by significant skyline residuals. Galaxy $i_0$ surface brightness is given.}\label{fig:halospec}
\end{center}
\end{figure}

Spectra were obtained as part of the SMEAGOL survey\footnote{http://sages.ucolick.org/surveys.html} on two separate dark nights: 2008 April 8 (hereafter Night 1) and 2009 March 23 (hereafter Night 2) using the DEep Imaging Multi-object Spectrograph (DEIMOS) on the Keck telescope. Three slit-masks were observed during Night 1 under good seeing conditions (FWHM $\sim 0.7$ arcsec) and 2 slit-masks were observed during Night 2, this time under variable seeing conditions ($0.8\lesssim$ FWHM $\lesssim1.3$ arcsec). The galaxy light data only for Night 1 were presented in both \citetalias{Proctor09} and \citetalias{Foster09}. In addition to the data used in \citetalias{Proctor09} and \citetalias{Foster09}, the final dataset used here includes roughly twice as many galaxy light spectra and the whole set of GC spectra. The 1200 l mm$^{-1}$ grating centered on 7800 \AA\space was used together with 1 arcsec slit width. This setup yields a resolution of $\Delta\lambda \sim 1.5$ \AA\space and allows for the coverage of the Calcium {\sc ii} Triplet (CaT) spectral region ($\sim$8400-8900 \AA). A total of 3 and 4 half-hour exposures were taken yielding a total exposure time of 1.5 and 2 hours per mask for Night 1 and 2, respectively. Fig. \ref{fig:slits} shows the positions of the slits that returned useful science spectra.

The DEIMOS data are reduced using the {\sc idl spec2d} data reduction pipeline provided online. Flat-fielding using internal flats, wavelength calibration using ArKrNeXe arc lamps, as well as the local sky subtraction are performed within the pipeline. The pipeline outputs the GC spectra with their corresponding fully propagated variance arrays as well as the subtracted background or `sky' spectra for each slit. Fig. \ref{fig:GCspec} shows example GC spectra for a range of signal-to-noise ratios.
 
We use the Stellar Kinematics with Multiple Slits (SKiMS) technique described in \citetalias{Proctor09} and \citetalias{Foster09} to extract the galaxy light spectra from these background spectra. Indeed, a background spectrum is essentially the sum of the pure sky and the galaxy light spectra. Therefore, if one has a good estimate of the sky's contribution to the background spectrum, it is possible to extract the galaxy light spectrum. For each mask, we obtain a high signal-to-noise estimate of the `pure' sky spectrum by summing and normalising several carefully selected background spectra at large radii (typically $\sim$6-7 $r_{e}$) where residual galaxy light is insignificant. This normalised sky spectrum is then scaled for each background spectrum using the sky scaling index defined in \citetalias{Foster09}\footnote{The sky scaling index was redefined in \citetalias{Foster09} in order to better avoid spectral features and differs slightly from that of \citetalias{Proctor09} \citepalias[see][for details]{Foster09}.} as the excess flux in the central passband with respect to the continuum level. The central passband of the sky scaling index (8605.0-8695.5 \AA) is measured between the blue (8478.0-8489.0 \AA) and red (8813.0-8822.0 \AA) continuum passbands. This scaled sky spectrum is subtracted from the background spectrum to obtain a galaxy light spectrum. As mentioned in \citetalias{Foster09}, this method yields a final continuum level accurate to 0.7 per cent of the noise in the skyline residuals in the sky index definition region. The majority ($\sim90$ per cent) of the amplitude of these skyline residuals are likely due to the intrinsic difficulties associated with non-local sky subtraction techniques such as variations of the sky spectrum across the DEIMOS field-of-view and with time. The additional $\sim 10$ per cent of the amplitude of the skyline residuals in our science galaxy spectra are caused by small variations in the wavelength solution across the mask. Fig. \ref{fig:halospec} shows example galaxy light spectra for a range of signal-to-noise ratios.

\section{Analysis and results}\label{sec:analysis}
In this section we describe how we extract the photometric, kinematic and stellar population information for both the galaxy light and the GC system. We also give an overview of the method used to fit the kinematics of NGC~4494. These results are interpreted and discussed in Section \ref{sec:discussion}.

	\subsection{Galaxy light}\label{sec:anal.GHL}
		
		\subsubsection{Stellar light profile}\label{sec:lum_profile}
\begin{figure}
\begin{center}
\includegraphics[width=84mm]{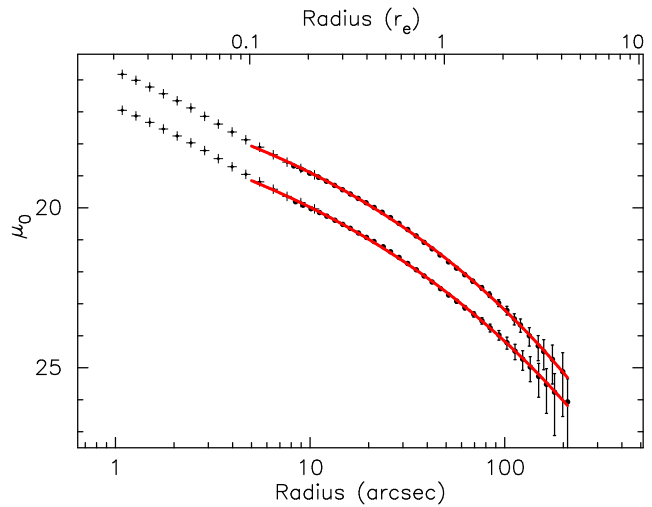} 
\caption{Surface brightness profiles for NGC~4494 as a function of intermediate radius (i.e., $\sqrt{ab}$ where $a$ and $b$ are the semi-major and minor axes, respectively). Filled circles are a compilation of our Subaru profiles and those from \citet{Lauer05}. Offsets were applied to the \citet{Lauer05} $V$ and $I$ profiles to match our $g$ and $i$ profiles, respectively. S\'ersic fits to the outer regions ($r>5$ arcsec) are shown as red solid lines.}\label{fig:lum_profile}
\end{center}
\end{figure}

We first quantify the stellar light distribution of NGC~4494. Figure \ref{fig:lum_profile} shows the surface brightness profile extracted as described in Section \ref{sec:phot} in the $g$- and $i$-bands together with literature values. We fit S\'ersic profiles \citep{Sersic63} to the surface brightness profile ($\mu(r)$) with geometric radius $r>5$ arcsec to avoid the inner disk as per \citetalias{Napolitano09}. In practice, we fit:
\begin{equation}\label{eq:Galsersic}
\mu(r)=\mu_e+\frac{2.5b_n}{\ln(10)}\left[(r/r_e)^{1/n}-1\right]
\end{equation}
\noindent where $b_n=1.9992n-0.3271$ \citep[i.e. eq. 6 of][]{Graham05} for the S\'ersic index ($n$), the effective radius ($r_e$) and the surface brightness at the effective radius ($\mu_e$). The resulting fits are shown in Fig. \ref{fig:lum_profile} and Table \ref{table:sersic}. We measure values between $48\le r_e\le55$ arcsec depending on the photometric filter used. Throughout this work we use the literature value of $r_e=49$ arcsec $\approx3.76$ kpc (Table \ref{table:props}) as it lies within the range of our measured values and is thus a good compromise.

\begin{table*}
\begin{center}
\begin{tabular}{ccccc}
\hline
Sample&$\mu_e \mid N_e$&$r_e \mid R_e$&$n$&$bg$\\
&(mag) $\mid$ (\#arcmin$^{-2}$)&(arcsec)&&(\#arcmin$^{-2}$)\\
\hline
galaxy g&$22.81\pm0.03$&$54.6\pm0.8$&$3.72\pm0.09$&---\\
galaxy r&$21.90\pm0.03$&$53.0\pm0.6$&$3.52\pm0.09$&---\\
galaxy i&$21.51\pm0.02$&$48.2\pm0.5$&$3.47\pm0.06$&---\\
all GCs&$5\pm1$&$100\pm10$&$1.7\pm0.5$&$0.28\pm0.02$\\
blue GCs&$2.2\pm0.8$&$138\pm24$&$1.8\pm0.7$&$0.22\pm0.03$\\
red GCs&$3.1\pm0.4$&$83\pm6$&$0.8\pm0.2$&$0.053\pm0.005$\\
\hline
\end{tabular}
\caption{Results of S\'ersic fits to the the galaxy surface brightness profile as a function of geometric radius (Eq. \ref{eq:Galsersic} fitted for $\mu_e$, $r_e$ and $n$) for points with $r>5$ arcsec. S\'ersic fits to the GC density profiles (Eq. \ref{eq:GCsersic} fitted for $N_e$, $R_e$, $n$ and $bg$) are also listed.}
\label{table:sersic}
\end{center}
\normalsize
\end{table*}	

		\subsubsection{Stellar kinematics}\label{sec:stelkin}

\begin{figure}
\begin{center}
\includegraphics[width=64mm]{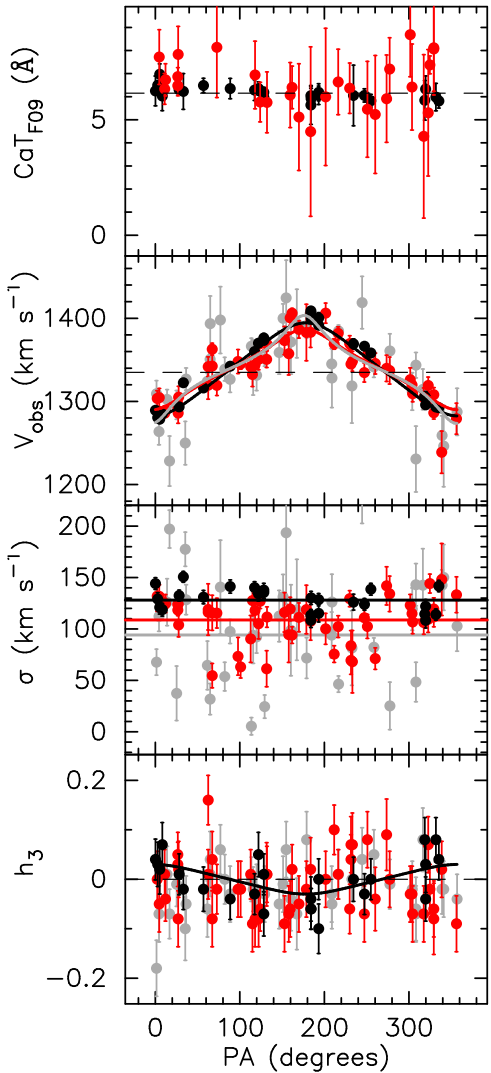}
\caption{Variation in $CaT_{F09}$ and velocity moments of the galaxy light as a function of PA for spectra with $r\le1r_e$ (black), $1r_e< r\le 2r_e$ (red), and $r>2r_e$ (gray). Kinemetry fits are shown with solid lines for the observed recession velocity ($V_{obs}$) and velocity dispersion in each radial bin, as well as for $h_3$ for spectra within 1$r_e$. Dashed lines represent the $CaT_{F09}$ saturation level (top panel), systemic velocity (second panel) and $h_3=0$ (bottom panel). The photometric position angle is $PA_{\rm phot}=173$ degrees. This figure is available in colour in the online version.}\label{fig:kinematics}
\end{center}
\end{figure}

	The stellar kinematics of NGC~4494 have been probed out to large radii in \citetalias{Proctor09} using the SKiMS method. Here our galaxy spectral sample is roughly twice that of \citetalias{Proctor09}. We measure the velocity moments (recession velocity, velocity dispersion and Gauss-Hermite coefficients $h_3$ and $h_4$) for all the galaxy light spectra using the {\sc pPXF} routine described in \citet{Cappellari04}. The {\sc pPXF} routine uses a set of 13 stellar templates to determine the best overall kinematic parameters and weighted combination of templates that minimises the residuals between the spectrum and the final resulting fit \citep[see][for more details]{Cappellari04}. The templates were observed using DEIMOS on the nights of the 2007 November 12-14 with a comparable instrumental setup. They cover a range of stellar sizes (11 giant and 2 dwarf stars) and span spectral types from F to early M, which dominate around this spectral region. The fitted spectral range is limited to 8450--8750 \AA\space for stability and regions heavily contaminated by skylines are not fitted (see Fig. \ref{fig:halospec}). Each fit is carefully inspected for quality control. Uncertainties on the measured velocity moments are estimated using Monte Carlo methods. For each spectrum, we randomly reshuffle the residuals between the best fit template and the original spectra in the wavelength region fitted by {\sc pPXF} before re-fitting. This is repeated 100 times for each individual spectrum. We use the standard deviation on the velocity moments for the 100 Monte Carlo realisations as our estimate of the random uncertainty.

Fig. \ref{fig:kinematics} shows the velocity moments as a function of position angle (PA) for our sample of spectra. Individual values can be found in Table \ref{table:halo}. The galaxy stars show clear major-axis rotation. The amplitude of the rotation is roughly constant all the way out to $>2r_e$ and shows a flattening (i.e., axis ratio $q_{\rm kin}=0.42\pm0.06$ is low) at large radii ($r\gtrsim2r_e$). The velocity dispersion decreases with radius. For the inner $h_3$ measurements, we find a clear trend with position angle although such a trend is not clearly visible beyond $1r_e$. The fourth velocity moment ($h_4$) has constant amplitude at all radii within the uncertainties.

In order to better understand the kinematic structure of NGC~4494, we use kinemetry to fit the model of an isotropic rotating ellipsoid (or inclined disk). Kinemetry is an analog to photometry where instead of fitting a model of the stellar light distribution, we fit a model of the kinematics  (see e.g. \citealt{Krajnovic06}; \citetalias{Proctor09}). In what follows, we generalise and improve on the technique presented in \citetalias{Proctor09} to obtain kinemetry fits to discrete and semi-discrete 2-dimensional data. \citetalias{Proctor09} have shown that kinemetry using this model provides a good fit to the inner regions probed by SAURON \citep{Emsellem04}. In contrast to \citetalias{Proctor09} where the data were binned in \emph{circular} annuli, here the data are binned in overlapping `radial' (semi-major axis) intervals that, together with the position angle and axis ratio, define \emph{`elliptical annuli'}. The $j^{{\rm th}}$ annulus (or bin) contains a subset of $N_j$ observed recession velocity data points ($V_{{\rm obs},i}$). We initially assume that the kinematic position angle and axis ratio of the galaxy coincide with the photometric values. The position angle and axis ratio of the $j^{{\rm th}}$ elliptical annulus are iteratively refined to match the kinematics as the algorithm fits for the kinematic position angle ($PA_{{\rm kin},j}$), axis ratio ($q_{{\rm kin},j}$) and the amplitude of the rotation ($V_{{\rm rot},j}$) using $\chi^2$ minimisation.

In practice, we first define an inner elliptical annulus ($j=1$) whose short semi-major axis length ($a_j$) is the distance to the closest data point and whose long semi-major axis length is $a_j + \Delta a_j$. We perform a $\chi^2$ minimisation of the data points contained within this elliptical annulus to the model. For the $j^{{\rm th}}$ annulus, the $\chi^2$ is computed using the following equation:
\begin{equation}\label{eq:Vobs1}
\chi^2_{V,j}=\sum^{i=N_j}_{i=1} \frac{1}{(\Delta V_{{\rm obs},i}')^2} \left( V_{{\rm obs},i}-V_{{\rm mod},i,j} \right)^2,
\end{equation}
where
\begin{equation}\label{eq:Vobs2}
V_{{\rm mod},i,j}=V_{{\rm sys}} \pm \frac{V_{{\rm rot},j}}{\sqrt{1+\left(\frac{\tan(PA_{i}-PA_{{\rm kin},j})}{q_{{\rm kin},j}}\right)^2}},
\end{equation}
and the ambivalent sign is positive (negative) if $(PA_{i}-PA_{{\rm kin},j})$ is in the first or fourth (second or third) quadrants. In Equations \ref {eq:Vobs1} and \ref {eq:Vobs2}, $PA_i$ and $\Delta V_{{\rm obs},i}'$ are the position angle and the uncertainty on the recession velocity measurement of the $i^{{\rm th}}$ data point, respectively. The systemic velocity of NGC~4494 ($V_{{\rm sys}}=1338.5$km s$^{-1}$) is measured by fitting the inner, highest signal-to-noise, data only prior to fitting the full dataset.

For each iteration within the $j^{{\rm th}}$ annulus, a new set of parameters ($V_{{\rm rot},j},PA_{{\rm kin},j},q_{{\rm kin},j}$) is found and the values of $PA_{{\rm kin},j}$ and $q_{{\rm kin},j}$ are used to update the shape and orientation of the elliptical annulus to be used in the next iteration ($a_j$ and $\Delta a_j$ remaining fixed). Therefore, by construction this changes the selected member data points in the $j^{{\rm th}}$ bin slightly for each iteration. After 15 iterations, we move out to the next `radial' bin (i.e. $(j+1)$) by increasing the length of the short semi-major axis such that $a_{j+1}=a_j+(\Delta a)/3$. The factor of 1/3 used here is chosen to increase the number of radial sub-samples or bins in order to smoothly define the kinemetry radial profile. The semi-major axis bin width ($\Delta a_{j+1}$) increases as a function of $a_{j+1}$ according to a de Vaucouleurs \citep{deVaucouleurs53} surface brightness profile (i.e. $\propto 1/(a_{j+1})^{1/4}$) in order to compensate for the lower signal-to-noise ratio at large galactocentric distances. With this setup, there are 15 overlapping bins, each containing $\sim25$ data points. This process continues until the radial extent of the data is covered.

\begin{figure*}
\begin{center}
\includegraphics[width=140mm]{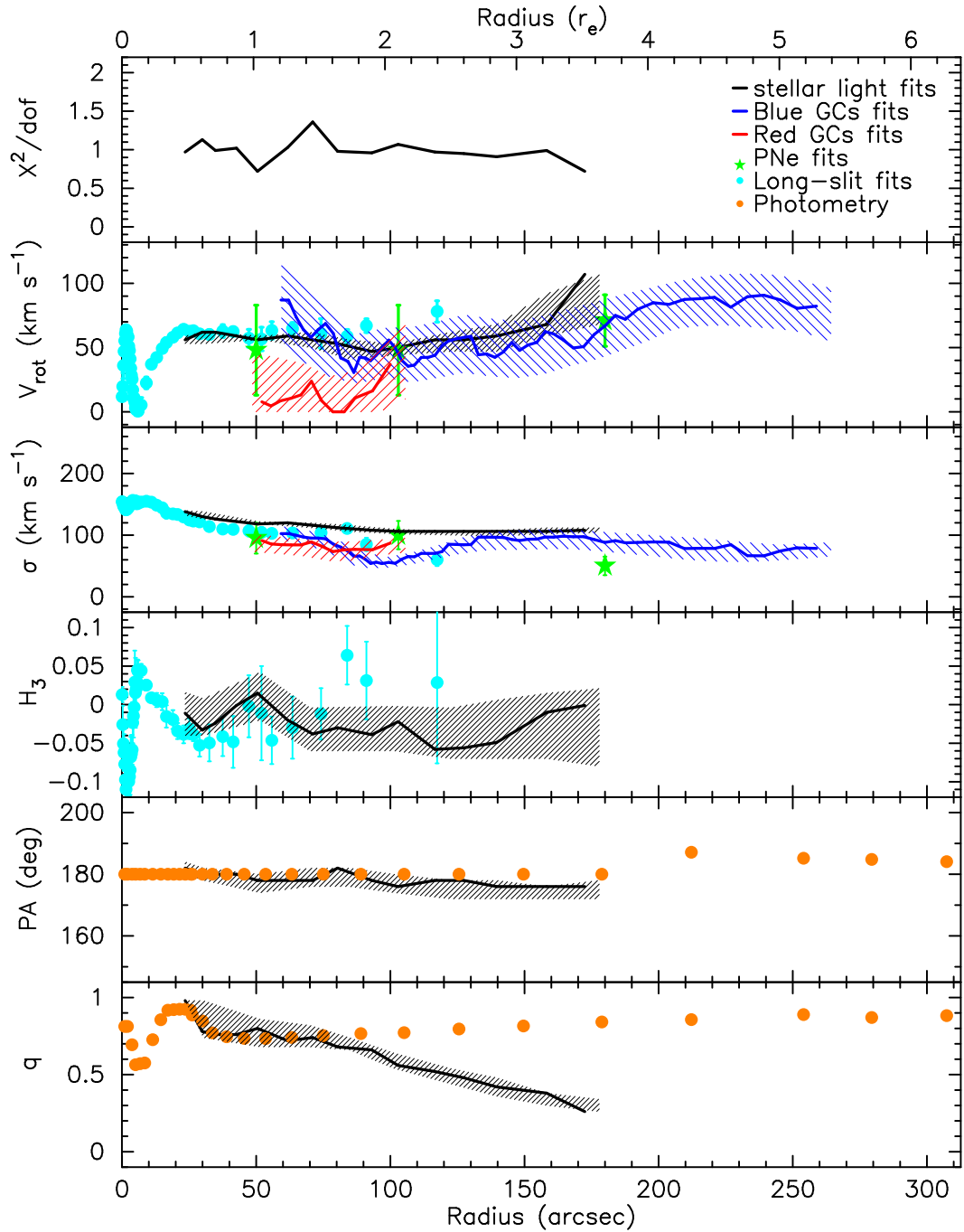}
\caption{Results of our kinemetry fits (solid lines) as a function of the major axis equivalent radius ($1r_e=49$ arcsec $\approx3.76$ kpc) with 68\% confidence intervals (hatched regions). Black, dark blue and red represent fits to the stellar light, the blue and red GCs, respectively. The top panel shows the value of the $\chi^2$ per degree-of-freedom for the fits of the first velocity moment only (i.e. $V_{rot}$). Light blue, green and orange filled symbols show results of \citetalias{Napolitano09} and \citet{Coccato09} from long-slit spectra, planetary nebulae (PNe) and galaxy light photometry, respectively. This figure is available in colour in the online version.}\label{fig:kinemetry}
\end{center}
\end{figure*}

%\begin{figure}
%\begin{center}
%\includegraphics[width=84mm]{??.eps}
%\caption{Reconstructed recession velocity (top panel) and velocity dispersion (lower panel) maps using the \citet{Coccato09} major-axis data in the inner parts (delimited by the black ellipses) and our radial kinemetry fits in the outer parts ($1r_e=49$ arcsec). Individual measurements shown as triangles. Note that the kinematically decoupled core in the central 19 arcsec is not modelled here for lack of required 2D spatial information. North is up and East is right. This figure is available in colour in the online version.}\label{fig:2dkinemetry}
%\end{center}
%\end{figure}

In order to obtain a $\chi^2/dof\approx1$ at all radii and particularly in the inner regions we add an uncertainty of 5 km s$^{-1}$ in quadrature to the Monte Carlo uncertainties (i.e., $\Delta V_{{\rm obs},i}'=\sqrt{(\Delta V_{{\rm obs},i})^2+(5 \:{\rm km \: s^{-1}})^2}$, see Table \ref{table:halo}). While the relative amplitudes of the uncertainties are crucial in order to obtain stable fits, this monotonic increase of the estimated uncertainties has a negligible effect on our fitted values. Nevertheless, it is an indication that either (1) there are velocity substructures at small galactocentric radii such that the model of a simple rotating isotropic ellipsoid is not appropriate or (2) the uncertainties are slightly underestimated. Indeed, the latter is likely since our quoted random uncertainties are obtained using Monte Carlo methods on our spectra and not independent measurements. For example, due to the finite (non-zero) length of the slits on the mask, their position angle, the effects of seeing, etc, independent measurements of the kinematics for the same ``position'' around the galaxy using either another instrument, setup or mask could yield slightly different values. For this reason, there could reasonably be some unaccounted for systematics of order $\sim 5$ km s$^{-1}$. Finally, two data points deviating from the best fit at the $>3\sigma$ level were excluded.

The kinemetry fits for the higher order velocity moments (velocity dispersion and $h_3$) are done in parallel in each annulus and using the values of $PA_{{\rm kin},j}$ and $q_{{\rm kin},j}$ obtained from the recession velocity kinemetry fits. In principle the position angle and axis ratio of these higher order velocity moments need not be equal to those of the recession velocity, but our data do not constrain these two parameters properly. Indeed, the velocity dispersion data in Fig. \ref{fig:kinemetry} show a hint of a dip along the minor axis ($PA\sim83$ and 263 degrees) for points beyond $\sim 1r_e$ (i.e., red and grey points), suggesting that the kinematic flattening (i.e., low $q_{\rm kin}$) is present in the velocity dispersion also. While this dip is suggestive, we are unable to reliably fit the axis ratio of the velocity dispersion explicitly with the current dataset and assume that of the recession velocity moment.

In practice, for the $j^{\rm th}$ bin we fit the velocity dispersion ($\sigma_j$) using $\chi^2$-minimization where 
\begin{equation}\label{eq:vd}
\chi^2_{\sigma,j}=\sum^{i=N_j}_{i=1} \left( \frac{\sigma_{{\rm obs},i}-\sigma_j}{\Delta\sigma_{{\rm obs},i}'} \right)^2.
\end{equation}
Eq. \ref{eq:vd} is different in form from Eq. \ref{eq:Vobs1} because the velocity dispersion is an even moment. In Eq. \ref{eq:vd}, $\sigma_{{\rm obs},i}$ and $\Delta\sigma_{{\rm obs},i}'=\sqrt{\Delta\sigma_{{\rm obs},i}^2+(8{\rm\:km\:s^{-1}})^2}$ are the measured recession velocity of the $i^{\rm th}$ spectrum and its associated random uncertainty ($\Delta\sigma_{{\rm obs},i}$) with an additional uncertainty of 8 km s$^{-1}$ added in quadrature in order to obtain a $\chi^2/dof\approx1$ as for the fits to $V_{{\rm obs}}$ above, respectively.

%Similarly, we minimize
%\begin{equation}\label{eq:H3_1}
%\chi^2_{H_{3},j}=\sum^{i=N_j}_{i=1} \frac{1}{(\Delta h_{3,{{\rm obs},i}})^2}(h_{3,{\rm obs},i}-h_{3,{{\rm mod},i,j}})^2,
%\end{equation}
%where
%\begin{equation}\label{eq:H3_2}
%h_{3,{{\rm mod},i,j}}=\pm\frac{H_{3,j}}{\sqrt{1+\left(\frac{\tan(PA_i-PA_{{\rm kin},j})}{q_{{\rm kin},j}} \right)^2}},
%\end{equation}
%to obtain $H_{3}$. The symbols $h_{3,{{\rm obs},i}}$ and $\Delta h_{3,{{\rm obs},i}}$ corresponds to the measured $h_3$ value and uncertainty of the $i^{\rm th}$ data point, respectively.

%Finally, we compute $H_4$ by minimising
%\begin{equation}\label{eq:H4}
%\chi^2_{H_{4},j}=\sum^{i=N_j}_{i=1} \left( \frac{h_{4,{{\rm obs},i}}-H_{4,j}}{\Delta h_{4,{{\rm obs},i}}} \right)^2,
%\end{equation}
%where $h_{4,{{\rm obs},i}}$ and $\Delta h_{4,{{\rm obs},i}}$ is the measured $h_4$ value and its uncertainty for the $i^{\rm th}$ slit.

A similar approach is applied to fit the \emph{amplitude} of the third ($H_3$) moment. Measurement uncertainties on individual velocity moments are propagated for the kinemetry fits using Monte Carlo methods. For each individual data point we have a measured uncertainty. Assuming that these uncertainties are Gaussian, we have a known distribution of possible measurements for each data point. Thus, for each point we randomly select from that distribution. Each time this is done for the entire dataset and we get a new set of possible measurements. The kinemetry algorithm described above is then applied to the new dataset. This exercise is repeated 100 times. From the resulting range of fits we determine the 68 per cent confidence interval of the best fit model allowed by the data.

We also use Monte Carlo methods to verify that kinemetry can be reliably applied to sparse data. We create a series of rotating ellipsoid models by varying the input rotational velocity $V_{\rm rot,in}=50$, 80 and 100 km s$^{-1}$, and kinematic axis ratio $q_{\rm kin,in}=0.5$, 0.8 and 1.0. For each model, we randomly sample 115 data points within $\sim3r_e$ and assign measurement errors consistent with our observed errors at that radius. This is repeated 25 times for each of the ($V_{\rm rot,in}$, $q_{\rm kin,in}$) combinations. We verify how well the input parameters are then recovered using kinemetry. In general, we find that the standard deviation of the output rotational velocity varies between $\sigma_{V_{\rm rot,out}}=5$ and 7 km s$^{-1}$, while $\sigma_{q_{\rm kin,out}}\sim0.1$ for all $q_{\rm kin,in}$. We find that the fits are less stable when both $V_{\rm rot,in}$ and $q_{\rm kin,in}$ are low. We do not find other ``unexpected'' systematics. We conclude that kinemetry can be applied to sparse data successfully provided that there is sufficient rotation.

The results of the kinemetry fits to the data are shown in Fig. \ref{fig:kinemetry}. In general, we find good agreement between the long-slit and planetary nebulae results of \citetalias{Napolitano09} and \citet{Coccato09} and our data from the SKiMS method. We have also tried fitting the PNe data from \citetalias{Napolitano09} using our method for discrete velocities (see Section \ref{sec:GCkin}) and find good general agreement with \citet{Coccato09} except around $2r_e\lesssim r\lesssim3r_e$ where the fits did not converge due to an apparently low rotation. The trends described above and found by \citetalias{Proctor09} are confirmed. We find sustained major-axis rotation at all radii consistent with both the results of \citetalias{Napolitano09} and \citet{Coccato09}. We do not find evidence for minor-axis rotation. There is a slight decrease in velocity dispersion that levels off with radius at a value of $\sim 100$km s$^{-1}$ beyond $\sim 2.3 r_e$ slightly at odds with the velocity dispersion estimated from planetary nebulae but consistent with that of the GCs (see Section \ref{sec:GCkin}). The amplitude of the third velocity moment levels off beyond $\sim1.8r_e$ at a value of $H_3\sim-0.03$.

\citet{Emsellem07} classified ETGs as slow- or fast-rotators according to a luminosity- and radius-weighted value of the angular momentum proxy $\lambda_R=V_{\rm rot}(r)/\sqrt{V_{\rm rot}(r)^2+\sigma(r)^2}$, where slow- and fast-rotators correspond to $\lambda_R < 0.1$ and 0.1, respectively. These classifications were made for radii $\le1r_e$, but \citetalias{Proctor09} demonstrated that the picture could change dramatically with more radially extended data. In the case of NGC~4494, we find that a local $\lambda_R\sim0.35$ (i.e. fast rotation) at virtually all probed radii, modulo a sharp dip around $0.1r_e$ due to the kinematically decoupled core. In \citet{Emsellem11}, NGC~4494 is also found to be a fast-rotator using a new classification method that takes into account the ellipticity of the isophotes.

Although not as pronounced as originally found in \citetalias{Proctor09}, a slight kinematic twist is visible beyond $\sim 2r_e$ such that the kinematic position angle is different from the photometric position angle at large radii. We measure a radial change in the axis ratio such that the velocity distribution becomes flattened (i.e., small $q_{\rm kin}$) as originally detected by \citetalias{Proctor09}. This behaviour is at odds with the measured isophotes that remain much rounder even at large radii with no sign of significant diskiness. We test the robustness of the flattening of $q_{\rm kin}$ at large radii using three methods. The first is the above described nominal method in which we allow $q_{{\rm kin}, j}$ to freely vary and iteratively define the bins accordingly. Secondly, similarly to the binning method used in \citetalias{Proctor09}, we let the fitted $q_{{\rm kin}, j}$ vary but fix the bin shapes to $q_{\rm phot}$, i.e. we do not let the kinematic axis ratio define the axis ratio of the bins. While this yields an overall larger $\chi^2/dof$ value, lower values of $q_{{\rm kin}, j}$ are still marginally preferred at large radii. Thirdly, we fix $q_{{\rm kin},j}=q_{\rm phot}$ for both the binning and the fitting and for all bins ($j$). This yields substantially larger $\chi^2/dof$ values than in the (first) nominal case. We conclude that the flattening of the kinematics is most robustly detected in the first method but is still marginally detected when the kinematic axis ratio is allowed to vary inside selection bins whose shape have fixed axis ratios. On the basis that the only self-consistent methods are the first and the third and given that the third method yields significantly higher $\chi^2/dof$ values than the first, we conclude that the former is preferable. We also tried removing the apparent ``outliers'' located around $PA\sim 0$ and 360 degrees with recession velocities below 1570 km s$^{-1}$ to see if they were causing the flattening and the results did not change significantly. We conclude that the kinematic flattening is robust (i.e., $q_{\rm kin}$ is indeed low). 

\begin{figure}
\begin{center}
\includegraphics[width=75mm]{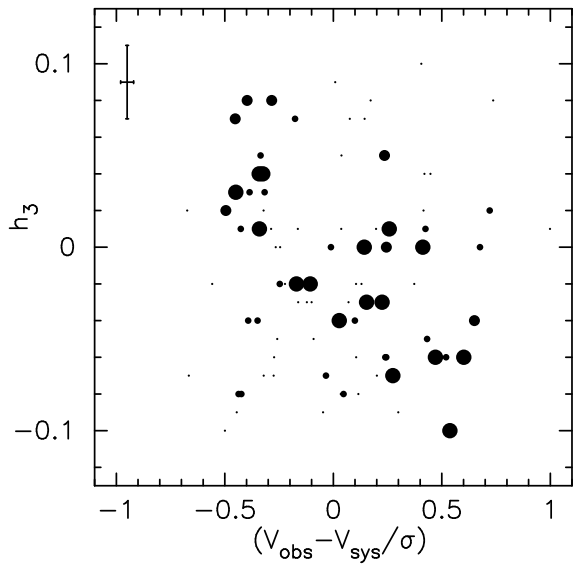}
\caption{Relationship between the higher order Gauss-Hermite moments $h_3$ and $(V_{\rm obs}-V_{\rm sys})/\sigma$. Point sizes are proportional to the signal-to-noise ratio of the spectra. The anti-correlation between $h_3$ and $(V_{\rm obs}-V_{\rm sys})/\sigma$ suggests a disk structure is present. Typical error bars shown in the upper left corner.}\label{fig:h3h4}
\end{center}
\end{figure}

Fig. \ref{fig:h3h4} shows the relationship between measured $h_3$ and $(V_{\rm obs}-V_{\rm sys})/\sigma$, which is a measure of the fraction of rotational over pressure support. The anti-correlation between $h_3$ and $(V_{\rm obs}-V_{\rm sys})/\sigma$ is indicative of a disk kinematic structure. This may be the source of the flattened kinematics.

		\subsubsection{Stellar colours and metallicities}\label{sec:starmet}
		
\begin{figure*}
\begin{center}
\includegraphics[width=175mm]{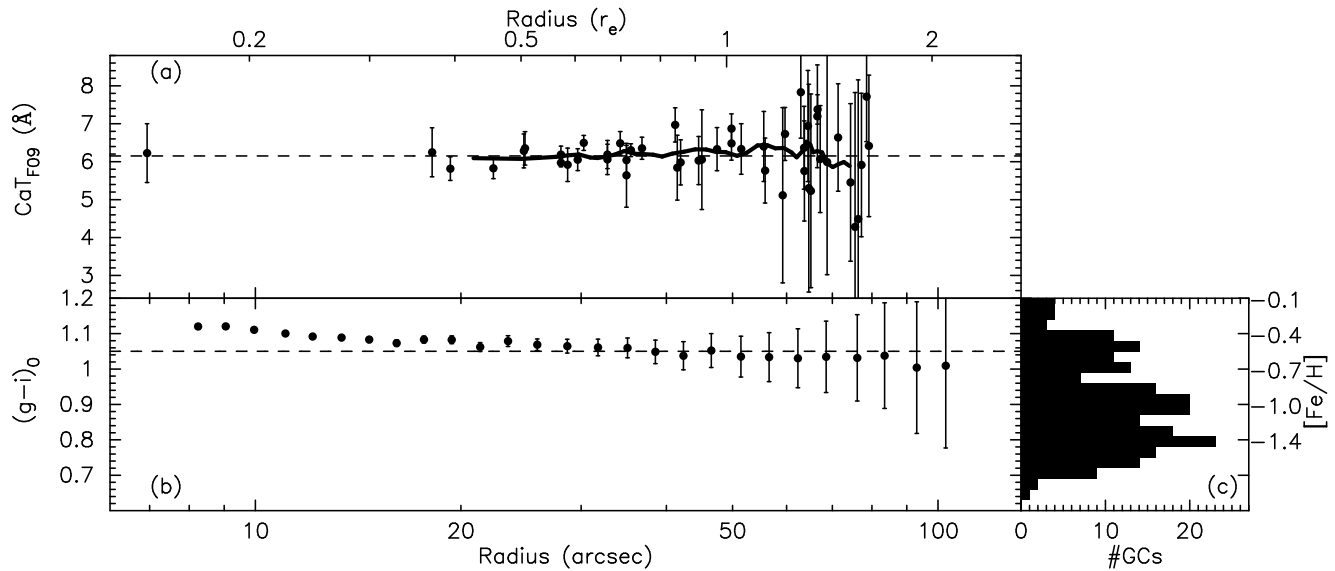}
\caption{Panels (a) and (b) show the $CaT_{\rm F09}$ index and $(g-i)_0$ colour gradients of the NGC~4494 stellar light, respectively. Thin dashed lines show the saturation limit predicted by the \citetalias{V03} SSP models (i.e., $CaT_{\rm F09}=6.18$\AA\space or $[Fe/H]\sim-0.5$ dex). Thick solid line in Panel (a) is a rolling average \citep[see e.g.,][]{Sawiloswsky07} using bins of 8 data points. Panel (c) is a colour histogram of the photometric sample of GCs with $i_0<23.5$ and $(g-i)$ conversion to metallicity as per \citet{Sinnott10}.}\label{fig:gradient}
\end{center}
\end{figure*}

We investigate the radial colour gradient of the stars in NGC~4494. By doing this, we obtain a \emph{rough} estimate of the radial metallicity gradient by assuming a fixed old GC-like age at all radii. In other words, we attribute any colour variation to changes in the metallicity (i.e. ignoring age effects). We convert $(g-i)_0$ colours into approximate metallicity using the empirical linear relationship derived by \citet{Sinnott10} for GCs in NGC~5128. This is shown in Fig. \ref{fig:gradient}. The galaxy stellar light is generally red and shows an overall colour/metallicity gradient. Because all the visible dust is contained within the inner 4 arcsec \citep{Lauer05}, we will assume that dust is not affecting our inferences on the metallicity at the radii probed here. Therefore, between 10 and 70 arcsec (or $\sim0.2r_e <r<1.4r_e$), the colour profile suggests a moderate metallicity variation with radius of $-0.17\pm 0.02$ dex per dex. Assuming a younger age would increase the colour-inferred metallicity. Beyond 2$r_e$ the photometric uncertainties start to dominate.

We use the CaT as a spectroscopic proxy of metallicity. The CaT has been employed as a metallicity indicator for resolved red giant stars \citep[e.g.,][]{Diaz89,Jorgensen92,Koch06}, integrated light spectra of Galactic \citep[e.g.,][]{Bica87,AZ88} and extragalactic \citep[][hereafter F10]{Foster10}\defcitealias{Foster10}{F10} GCs as well as galaxies (e.g., \citealt{Cenarro08a}; \citetalias{Foster09}). We compute the CaT index value for each galaxy spectrum using the index definition from \citetalias{Foster09} (i.e. $CaT_{\rm F09}$). The $CaT_{\rm F09}$ is corrected for velocity dispersion broadening and converted into $[Fe/H]$ using the single stellar population (SSP) models of \citet[hereafter V03]{V03} following \citetalias{Foster09}. Once again, we emphasise that the inferred galaxy starlight metallicities are uncertain due to the above caveats. Fig. \ref{fig:gradient} shows the CaT index as a function of the semi-major axis equivalent radius (as per equations 3 and 4 of \citetalias{Foster09}). The $CaT_{\rm F09}$ gradient is found to be essentially flat at all probed radii indicating an undetectably small radial change in metallicity.

The \citetalias{V03} SSP models predict that the CaT features saturate at a metallicity of roughly ${\rm [Fe/H]}\approx-0.5$ such that there is a maximum allowed $CaT_{\rm F09}$ value of 6.18 \AA\footnote{This is different from the saturation value for ${\rm CaT_{F10}}$ because $CaT_{\rm F09}$ is an altogether different index. Therefore, the absolute values of the two CaT indices used in this work for GC and galaxy light spectra cannot be directly compared.}. This theoretical behaviour has not been confirmed observationally. Our galaxy $CaT_{\rm F09}$ data are consistent with a saturation around 6.18 \AA. Because our data scatter about this limit we cannot confidently use the \citetalias{V03} models to convert our $CaT_{\rm F09}$ values into metallicity. It is possible however that, as the colours suggest, the metallicity variation in the radial range probed is small (only $\Delta[Fe/H]\sim0.2$ dex) and thus a variation in $CaT_{\rm F09}$ may be hard to detect within the uncertainties. Indeed, between 20 and 80 arcsec, where the vast majority of our CaT data lies, both the colours and $CaT_{\rm F09}$ values are consistent with the predicted saturation limit.

\begin{table}
\begin{center}
\begin{tabular}{ccc}
\hline
Passband&$CaT_{F10}$&$CaT_{F09}$\\
&(\AA)&(\AA)\\
\hline
Ca1&8490.0-8506.0&8483.0-8513.0\\
Ca2&8532.0-8552.0&8527.0-8557.0\\
Ca3&8653.0-8671.0&8647.0-8677.0\\
\hline
&&8474.0-8483.0\\
&&8514.0-8526.0\\
Shared continuum&&8563.0-8577.0\\
&&8619.0-8642.0\\
&&8680.0-8705.0\\
\hline
\end{tabular}
\caption{CaT index definition for galaxy stellar light spectra ($CaT_{\rm F09}$) and GC spectra ($CaT_{\rm F10}$).}
\label{table:CaT}
\end{center}
\end{table}	

	\subsection{Globular cluster system}\label{sec:anal.GCS}
	
		\subsubsection{GC colours and candidate selection}\label{sec:selection}		
\begin{figure}
\begin{center}
\includegraphics[width=84mm]{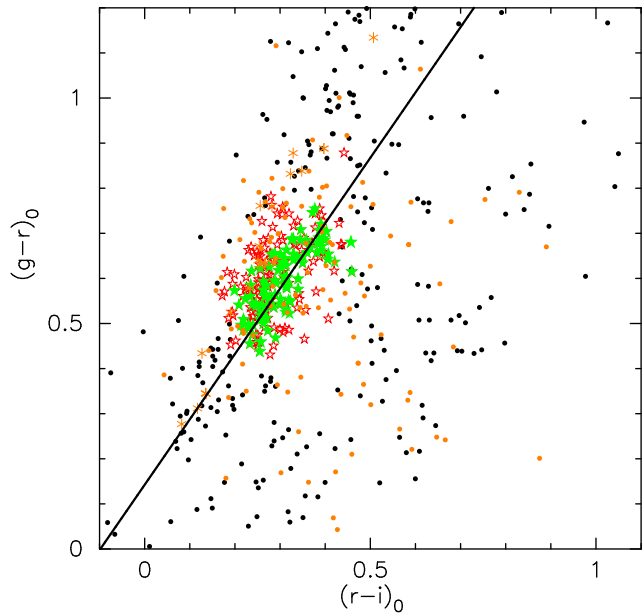}
\caption{Visual representation of our applied colour-colour selection for objects with galactocentric radii $<8$ arcmin brighter than $i_0=23.5$. Filled green star symbols are spectroscopically confirmed GCs while hollow red stars show our photometrically selected GC candidates. Solid line is the inferred GC sequence (i.e., $(g-r)_0=1.450(r-i)_0+0.142$). Spectroscopically confirmed stars (orange asterix) and emission line galaxies (orange filled circles) observed either as contaminants or fillers are shown. This figure is available in colour in the online version.}\label{fig:cc}
\end{center}
\end{figure}

As described in Section \ref{sec:phot}, we have a combination of both HST/WFPC2 and Subaru/Suprime-Cam imaging available. Various selections are applied to identified point sources in order to avoid contamination by foreground stars and background galaxies. We first apply an upper size cut in both the HST and the Subaru images\footnote{The size cut based on Subaru is only applied to objects beyond galactocentric radii of 0.5 arcmin because of crowding issues.} in order to remove clearly extended objects such as background galaxies. We also apply a spatial cut removing all objects beyond 8 arcmin ($\sim 10 r_e$) from the galactic centre to avoid further contamination that dominates beyond that radius (see \ref{sec:GCdensprof}). Our colour selections are as follows: (1) whenever HST imaging is available, GC candidates are selected based on $(V-I)_0$ colours only. (2) If HST imaging is not available, the selection is based on a colour-colour cut in $(r-i)_0$ vs $(g-r)_0$ space from Subaru photometry (see Fig. \ref{fig:cc}). 
%We fit for the linear bisector \citep{Feigelson92} in colour-colour space using the spectroscopically confirmed GCs (i.e., $(g-r)_0=1.450\times(r-i)_0+0.142$) and allow for three times the intrinsic scatter about that line ($\sigma=0.05$) for objects with $0.60\le(g-i)_0\le1.20$, $0.05\le(g-r)_0\le0.75$ and $0.20\le(r-i)_0\le0.40$. Objects whose photometric uncertainties are sufficiently large to overlap with the above selection ranges are also kept as GC candidates. 
The final master GC catalogue of Subaru and HST photometry includes 431 selected GC candidates brighter than $i_0=24$. This catalogue is available online (also see Table \ref{table:gccat}).

\begin{table*}
\begin{center}
\caption{Catalog of photometrically selected GC candidates with $i_0<24$. Columns 1 and 2 give the position in right ascencion and declination (J2000), respectively. Columns 3 to 7 are the Subaru/Suprime-Cam and HST/WFPC2 photometry. The full table is available in the online version.}
\begin{tabular}{ccccccc}
\hline
$\alpha$&$\delta$&$g_0$&$r_0$&$i_0$&$V_0$&$I_0$\\
(hh:mm:ss)&(hh:mm:ss)&(mag)&(mag)&(mag)&(mag)&(mag)\\
(1)&(2)&(3)&(4)&(5)&(6)&(7)\\
\hline
 12:31:23.70& 25:40:12.90&24.072$\pm$0.023&23.359$\pm$0.020&22.981$\pm$0.019&---&---\\
 12:31:07.44& 25:40:15.96&24.362$\pm$0.026&23.672$\pm$0.022&23.319$\pm$0.021&---&---\\
 12:31:38.83& 25:40:17.01&24.922$\pm$0.032&24.271$\pm$0.030&23.830$\pm$0.028&---&---\\
 12:31:08.79& 25:40:27.99&23.805$\pm$0.022&23.237$\pm$0.018&22.991$\pm$0.020&---&---\\
 12:31:18.50& 25:40:29.10&23.377$\pm$0.020&22.823$\pm$0.016&22.562$\pm$0.018&---&---\\
 12:31:36.43& 25:40:34.57&22.326$\pm$0.019&21.864$\pm$0.015&21.662$\pm$0.016&---&---\\
 ...&...&...&...&...&...&...\\
\hline
\end{tabular}\label{table:gccat}
\end{center}
\end{table*}	

Selecting bright objects in common between the HST and Subaru images beyond a galactocentric radius of 0.5 arcmin to avoid Subaru image artefacts, we obtain the following colour conversion:
\begin{equation}
(g-i)_0=1.23\times(V-I)_0-0.25
\end{equation}
with an rms$=0.035$ using a linear bisector fit \citep{Feigelson92}. We use this conversion to convert HST $(V-I)_0$ colours into $(g-i)_0$ to merge with the Subaru photometry.

\begin{figure}
\begin{center}
\includegraphics[width=84mm]{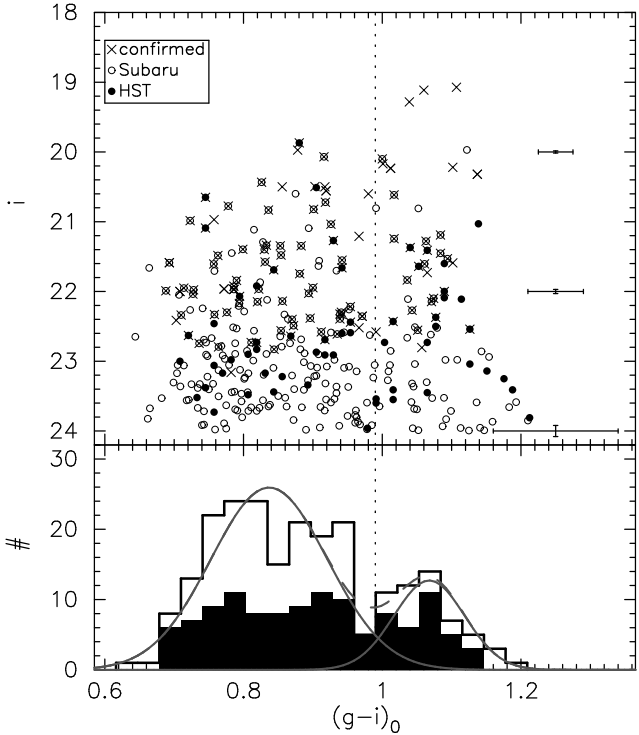} 
\caption{CMDs (top panel) for our GC candidates beyond 0.5 arcmin from the galactic centre selected based on Subaru (hollow circles) and HST (filled circles) imaging. All spectroscopically confirmed GCs around NGC~4494 are shown as crosses. Typical Subaru error bars are shown on the right hand side of the top panel. Lower panel shows histograms for our candidates brighter than $i_0=23.5$ (hollow) with KMM fits for each/the sum of the two subpopulations as solid/dashed grey line(s) and confirmed (filled histogram) GCs. The colour distribution of the confirmed GCs is representative of that of the candidates. Dotted line represents our fiducial colour split.}\label{fig:cmd}
\end{center}
\end{figure}

The final colour magnitude diagram (CMD) for all our GC candidates with $i_0<24$ is shown in Fig. \ref{fig:cmd}. Three bright objects ($i_0<19.4$ mag) have recession velocities consistent with that of NGC~4494 but their absolute luminosity ($M_i<-10.9$ mag) suggests they fit within the definition of ultra-compact dwarfs (UCDs), or equivalently, dwarf-globular transition objects (DGTOs, see Section {\ref{sec:UCD}}). All three UCDs are centrally located (within 2 arcmin or $\sim4r_e$ of NGC~4494's centre). Other bright objects ($i_0<19.4$ mag) with colours consistent with the NGC~4494 GCs are uniformly distributed across the Subaru/Suprime-Cam field-of-view (19$\times$15 arcmin) suggesting that they are likely contaminant foreground stars. For this reason we apply a bright magnitude cut at $i_0=19.4$.

We check for the usual colour bimodality within our GC candidates in the combined HST and Subaru $(g-i)_0$ colour distributions. We apply the KMM \citep{Ashman94} test for the heteroscedastic (unequal widths) case to our GC candidates with $i_0<23.5$. The colour histogram and fit results are shown in Fig. \ref{fig:cmd}. The returned p-value is $<0.0001$, suggesting that a bimodal colour distribution is strongly favoured over a unimodal one at the $>99.99$ per cent level. The peaks for the blue and red GCs are at $(g-i)_0=0.84$ and 1.07, with widths of $\sigma=0.084$ and 0.051, respectively. For comparison, \citet{Larsen01} find peaks at $(V-I)_0=0.90$ and 1.10, corresponding to $(g-i)_0=0.86$ and 1.10, for the blue and red subpopulations when using the homoscedastic test on the HST data only. These small discrepancies are likely due to the intrinsic differences between the datasets (HST vs Subaru), possible intrinsic radial colour gradients \citep[e.g.,][]{Harris09a} and methods (homoscedastic vs heteroscedastic). \citet{Larsen01} also infer a relatively equal number of GCs in each subpopulation. We choose to apply a nominal colour split at $(g-i)_0=0.99$, corresponding to $(V-I)_0\sim1.01$ based on the KMM results in order to separate the blue and red GCs. Using the \citet{Sinnott10} conversion from $(g-i)_0$-colour to metallicity suggests that our colour split corresponds to a metallicity of $[Fe/H]\sim-0.68$ dex.

		\subsubsection{GC spatial distribution}\label{sec:GCdensprof}
\begin{figure}
\begin{center}
\includegraphics[width=84mm]{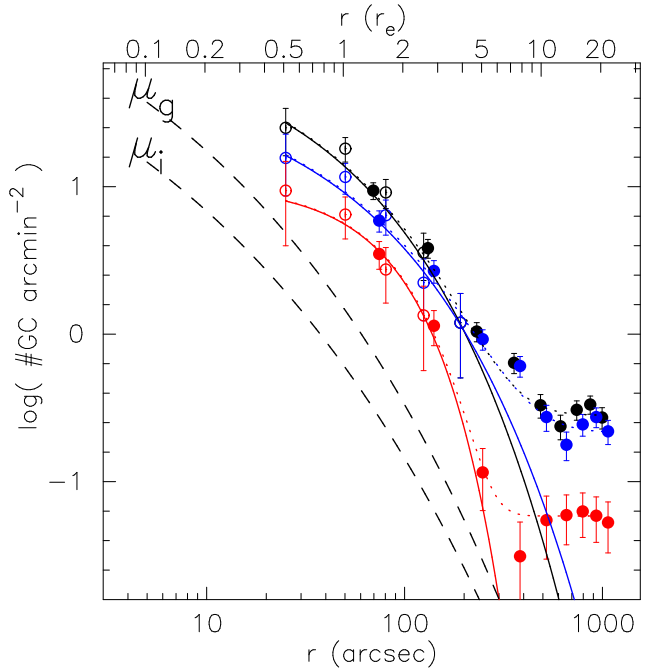}
\caption{Surface density profiles of the NGC~4494 GC system. All, red and blue subpopulations are shown in black, red and blue, respectively. Open and filled symbols show the HST and Subaru data, respectively. S\'ersic fits with and without background/contamination constant (see text) are shown as dotted and solid lines, respectively. Also shown are scaled and offset S\'ersic profile fits to the galaxy surface brightness profiles in the $g$- and $i$-bands ($\mu$) as labelled (see Section \ref{sec:lum_profile} and Fig. \ref{fig:lum_profile}). This figure is available in colour in the online version.}\label{fig:GCdens}
\end{center}
\end{figure}

We construct projected surface density profiles of the NGC~4494 GC system candidates.  Independent photometric selection criteria (as described in Section \ref{sec:selection}) are applied to both the HST and Subaru datasets to generate GC catalogues for this analysis. Between 1 and 4 arcmin, there are reliable surface density data for both Subaru and HST GC catalogues. Within the uncertainties the Subaru profile agrees with the HST profile, as shown in Figure \ref{fig:GCdens}.

We fit the GC density profile ($N(r)$) with a S\'ersic profile \citep{Sersic63} similar to that commonly done for galaxy surface density profiles and recently extended to GC systems \citep{Peng11}. We fit the following variation of eq. 1 from \citet{Graham05}:
\begin{equation}\label{eq:GCsersic}
N(r)=N_e\exp\left(-b_n\left[(r/R_{e})^{1/n}-1\right)\right]+bg,
\end{equation}
\noindent where $b_n=1.9992n-0.3271$. Free parameters recovered are the S\'ersic index (n) of the GC system, which is a measure of the steepness of the profile, the effective radius of the GC system ($R_e$), which gives us a measure of the extent of the GC system, the surface density at that radius ($N_e$) and the background or contamination level ($bg$). The results of the fits are shown in Fig. \ref{fig:GCdens} and Table \ref{table:sersic}. For comparison with previous studies, we also fit a Power-law profile (i.e., $N(r)\propto r^\alpha$). We obtain slopes of $\alpha=-1.7\pm0.2$, $-1.8\pm0.2$ and $-2.2\pm0.3$ for all, blue and red GCs, respectively. As with other galaxies \citep[e.g.,][]{Forbes98a,Dirsch03,Bassino06}, the red GCs are more centrally concentrated than the blue GCs. This is also inferred from their respective effective radii ($R_{e,{\rm blue}}=138$ arcsec and $R_{e,{\rm red}}=83$ arcsec).

We use 500 Monte Carlo realisations to estimate the total number of GCs based on the best fit S\'ersic parameters and their covariance matrix by separately integrating the density profiles for all, the blue and the red GCs to infinity. We also explicitly account for completeness and contamination. This yields an estimated number of GCs of $392\pm49$, $324\pm74$ and $125\pm10$ for all, red and blue GCs, respectively. Therefore, assuming $m_V=9.70$ (RC3) as total galaxy magnitude and using other variables as per Table \ref{table:props}, the specific GC frequency $S_N$ \citep{Harris81} is $1.2\pm0.3$, while $T_N$ \citep{Zepf93} is $4\pm1$. This is somewhat low for galaxies of similar stellar mass but still within the observed range \citep[see][]{Spitler08b,Peng08}.

It has recently been shown that the correlation between total number of GCs and the central black-hole mass \citep{Spitler09} is tighter than other previously observed correlations with black-hole mass \citep[see][]{Burkert10,Harris11,Snyder11}. Therefore, using equation 3 of \citet{Harris11}, we estimate that the mass of the central black-hole in NGC~4494 is $(1.6\pm0.2)\times10^8M_{\odot}$.

		\subsubsection{GC kinematics}\label{sec:GCkin}
\defcitealias{Cote03}{C03}		
\defcitealias{Hwang08}{H08}
\defcitealias{Cote01}{C01}
\defcitealias{Schuberth10}{S10}
\defcitealias{Romanowsky09}{R09}
\defcitealias{Bergond06}{B06}
\defcitealias{Lee10}{L10}
\defcitealias{Woodley10a}{W10a}

\begin{table*}
\scriptsize
\begin{center}
\begin{tabular}{cccccccccccc}
\hline
Galaxy&$V_{\rm rot,all}$&$PA_{\rm kin,all}$&$\sigma_{\rm all}$&$V_{\rm rot,blue}$&$PA_{\rm kin,blue}$&$\sigma_{\rm blue}$&$V_{\rm rot,red}$&$PA_{\rm kin,red}$&$\sigma_{\rm red}$&$N_{GC}$&Reference\\
&(km s$^{-1}$)&(deg)&(km s$^{-1}$)&(km s$^{-1}$)&(deg)&(km s$^{-1}$)&(km s$^{-1}$)&(deg)&(km s$^{-1}$)&&\\
(1)&(2)&(3)&(4)&(5)&(6)&(7)&(8)&(9)&(10)&(11)&(12)\\
\hline
M49&$53^{+52}_{-25}$&$105^{+45}_{-45}$&$312^{+27}_{-8}$&$93^{+69}_{-37}$&$100^{+37}_{-40}$&$342^{+33}_{-18}$&$12^{+76}_{-74}$&$195^{+56}_{-58}$&$265^{+34}_{-13}$&263&\citetalias{Cote03}\\
M60&$141^{+50}_{-38}$&$225^{+12}_{-14}$&$234^{+13}_{-14}$&$130^{+62}_{-51}$&$218^{+16}_{-23}$&$223^{+13}_{-16}$&$171^{+58}_{-46}$&$237^{+18}_{-19}$&$258^{+21}_{-31}$&121&\citetalias{Hwang08}\\
M87&$169^{+42}_{-97}$&$66\pm35$&$384^{+27}_{-32}$&$172^{+51}_{-108}$&$59\pm52$&$397^{+37}_{-46}$&$160^{+120}_{-99}$&$76\pm45$&$364^{+49}_{-52}$&278&\citetalias{Cote01}\\
NGC 1399&---&---&---&$110\pm53$ &$130\pm24$& $333\pm16$ &$61\pm35$&$154\pm33$&$239\pm11$ &$\sim670$&\citetalias{Schuberth10}\\
NGC 1407&$86^{+32}_{-39}$&$46^{+22}_{-21}$&$241^{+14}_{-12}$&$87^{+37}_{-57}$&$91^{+36}_{-32}$&$234^{+21}_{-16}$&$104^{+43}_{-53}$&$29^{+23}_{-28}$&$247^{+27}_{-17}$&156&\citetalias{Romanowsky09}\\
NGC 3379&$<100$&---&$169\pm20$&---&---&---&---&---&---&30&\citetalias{Bergond06}\\
NGC 4494&$56^{+22}_{-5}$&$147^{+12}_{-18}$&$91^{+5}_{-9}$&$62^{+16}_{-11}$&$170^{+16}_{-14}$&$90^{+6}_{-9}$&$35^{+21}_{-20}$&(173)\footnotemark&$92^{+8}_{-21}$&109&this work\\
NGC 4636&$37^{+32}_{-30}$&$174^{+73}_{-48}$&$225^{+12}_{-9}$&$27^{+34}_{-24}$&$0^{+146}_{-144}$&$251^{+18}_{-12}$&$68^{+48}_{-35}$&$178^{+53}_{-34}$&$203^{+12}_{-13}$&238&\citetalias{Lee10}\\
NGC 5128&$33\pm10$&$185\pm15$&---&$26\pm15$&$177\pm28$&$149\pm4$&$43\pm15$&$196\pm17$&$156\pm4$&564&\citetalias{Woodley10a}\\
\hline
\end{tabular}
\caption{Compilation of salient GC kinematic properties in the literature for selected large ETGs (column 1). Rotation amplitudes for all, blue and red GCs are given in columns 2, 5 and 8, respectively. The position angle of the rotation for all, blue and red GCs are given in columns 3, 6 and 9, respectively. Photometric position angles for each galaxy can be found in Table \ref{table:props} The velocity dispersion for all, blue and red GCs are shown in columns 4, 7 and 10, respectively. The number of GCs in the kinematic sample for each study is shown in column 11. References in column 12 correspond to \citet[][C03]{Cote03}, \citet[][H08]{Hwang08}, \citet[][C01]{Cote01}, \citet[][S10]{Schuberth10}, \citet[][R09]{Romanowsky09}, \citet[][B06]{Bergond06}, \citet[][L10]{Lee10} and \citet[][W10a]{Woodley10a}.}
\label{table:litGCkin}
\end{center}
\end{table*}
\footnotetext{We fix $PA_{\rm kin}=PA_{\rm phot}=173$ degrees to get a better handle on the red GC kinematics since the rotation is low (see text).}

For the candidate GCs that have available spectra with sufficient signal-to-noise ratio, we first measure their recession velocity. We use the {\sc iraf} task {\sc rv.fxcor} to cross-correlate the spectra with the 13 stellar templates described in Section \ref{sec:stelkin}. Thus, for each GC we have 13 measured recession velocities together with output uncertainties from the {\sc fxcor} routine. The recession velocities reported in Table \ref{table:GC} are the average measured recession velocities using all 13 template cross-correlation results. The quoted uncertainties for the GC recession velocities are the maximum between 5 km s$^{-1}$ or the average output uncertainties given by {\sc fxcor}, which correspond to the average width of the cross-correlation peaks, added in quadrature to the standard deviation among the templates, which is an estimate of the systematics. Some of our spectra cover sufficiently blue wavelengths to use the H${\rm \alpha}$ feature at 6563\AA\space to confirm the measured recession velocity. Some of our spectra were from Galactic stars and background emission line galaxies. Table \ref{table:contcat} shows the position of those contaminants and their recession velocity.

\begin{table*}
\begin{center}
\caption{List of spectroscopically identified contaminants and non-globular cluster fillers. Columns 1 and 2 give the position in right ascension and declination (J2000), respectively. Columns 3, 4 and 5 are the Subaru/Suprime-Cam photometry. Measured recession velocities, when available, can be found in column 6. Column 7 lists the type of contaminant. The full table is available online.}
\begin{tabular}{ccccccc}
\hline
$\alpha$&$\delta$&$g_0$&$r_0$&$i_0$&$V_{\rm obs}$&type\\
(hh:mm:ss)&(hh:mm:ss)&(mag)&(mag)&(mag)&(km s$^{-1}$&\\
(1)&(2)&(3)&(4)&(5)&(6)&(7)\\
\hline
 12:31:22.33& 25:49:24.25&23.987$\pm$0.029&23.312$\pm$0.020&22.727$\pm$0.020&---&galaxy\\
 12:31:25.91& 25:42:55.93&23.960$\pm$0.023&23.512$\pm$0.022&22.828$\pm$0.020&---&galaxy\\
 12:31:09.93& 25:52:22.29&23.570$\pm$0.021&22.813$\pm$0.017&22.507$\pm$0.018&---&galaxy\\
 12:31:09.95& 25:51:11.12&24.591$\pm$0.027&24.331$\pm$0.035&23.990$\pm$0.031&---&galaxy\\
 12:31:31.08& 25:40:47.79&23.337$\pm$0.020&22.927$\pm$0.017&22.710$\pm$0.019&---&galaxy\\
 12:31:26.10& 25:40:45.25&23.602$\pm$0.021&23.360$\pm$0.018&22.694$\pm$0.018&---&galaxy\\
 ...& ...&...&...&...&...&...\\
\hline
\end{tabular}
\label{table:contcat}
\end{center}
\end{table*} 

The spatial distribution of kinematically confirmed GCs is shown in Fig. \ref{fig:slits}. We look for rotation in the GC system around NGC~4494. We extend the kinemetry method described in Section \ref{sec:stelkin} to discrete velocities and perform rolling or moving radial fits for the amplitude of the rotation ($V_{{\rm rot}}$), the velocity dispersion ($\sigma$) and the kinematic position angle ($PA_{{\rm kin}}$) simultaneously. Rolling fits are similar in principle to rolling or moving averages \citep{Sawiloswsky07} and were also used in \citetalias{Proctor09} \citep[see also][who use a very similar method]{Kissler-Patig98}. They are performed by first using the inner $N_{\rm bin}$ GCs to fit the kinemetry, then removing the innermost point and adding the next further point to refit the kinemetry, and so on, until all GCs have been exhausted. We use bins of 25 and 20 for the blue and red GCs, respectively. Gaussian line-of-sight velocity distributions are assumed. In practice, for the $j^{\rm th}$ radial bin we minimise the likelihood ratio ($\Lambda_j$):

\begin{equation}\label{eq:GC_LR}
\Lambda_{j}\propto\sum^{i=N_j}_{i=1} \left[\frac{(V_{{\rm obs},i}-V_{{\rm mod},i,j})^2}{(\sigma_j^2+(\Delta V_{{\rm obs},i})^2)}+\ln (\sigma_j^2+(\Delta V_{{\rm obs},i})^2)\right],
\end{equation}
where $V_{{\rm mod},i,j}$ is again given by Eq. \ref{eq:Vobs2} with $q_{{\rm kin},j}=q_{{\rm phot}}=0.87$. Symbols $PA_i$, $V_{{\rm obs},i}$ and $\Delta V_{{\rm obs},i}$ are the position angle, recession velocity and uncertainty on the recession velocity for the $i^{{\rm th}}$ GC, respectively. We assume that the kinematic axis ratio of the GC system is equal to the photometric axis ratio of the galaxy light because the GC kinematic data do not constrain the kinematic axis (i.e., $q_{{\rm kin},j}$) ratio well.

\begin{figure*}
\begin{center}
\includegraphics[width=178mm]{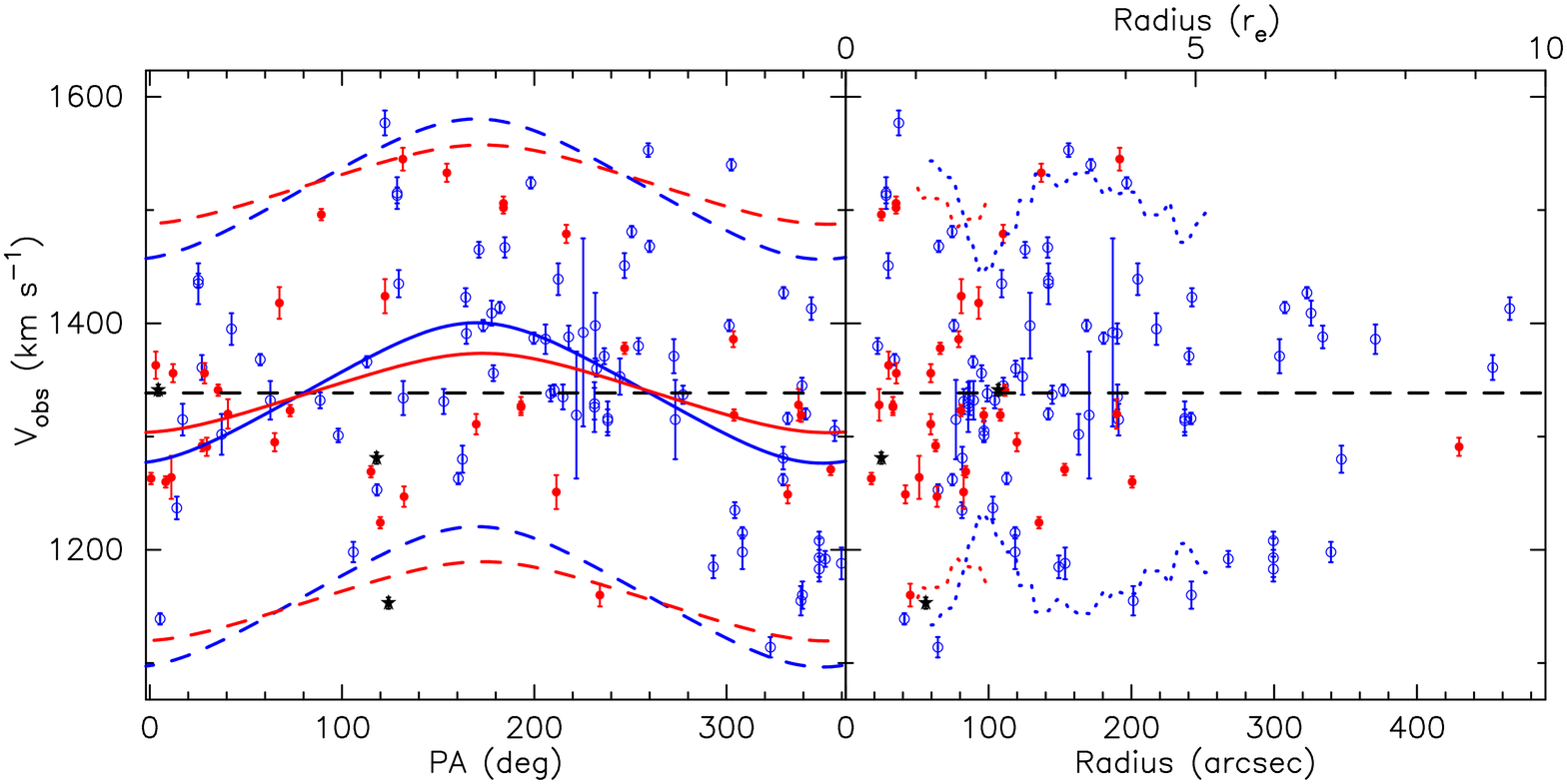}
\caption{GC velocity distribution of the blue (hollow blue symbols), red (filled red symbols) GCs and UCDs (star symbols) recession velocities as a function of position angle (left panel) and galactocentric radius (right panel). Solid and dashed blue/red lines show the kinematic fit results and $\pm2\sigma$, respectively, for the blue/red GC subpopulation. Blue/red dotted lines show 2$\times$ the velocity dispersion (corrected for rotation) intervals for the blue/red GC subpopulation. Black dashed line represents NGC~4494's systemic velocity. The photometric position angle is $PA_{\rm kin}=173$ deg. This figure is available in colour in the online version.}\label{fig:GCkin}
\end{center}
\end{figure*}

Uncertainties on the fits to the GC kinematics are obtained using a bootstrapping method similar to that used by \citet{Cote01}. Basically, we obtain 1000 ``mock'' GC kinematic samples of 109 GCs (our actual sample size) by sampling with replacement from our measured distribution and fit each mock sample. The quoted uncertainties are the 68\% confidence interval inferred from the mock fits.

We also notice that this method tends to enhance the estimated rotation value when the position angle varies as a free parameter. We used Monte Carlo methods to quantify this bias. We find that $V_{\rm rot}$ estimated using the above method on a sample of the same bin size and known slow rotation ($V_{\rm rot}<10$ km s$^{-1}$) can be enhanced by as much as 20 km s$^{-1}$ when $PA_{\rm kin}$ is allowed to vary freely. This bias sharply reduces to $\lesssim5$ km s$^{-1}$ for known input $V_{\rm rot}>50$ km s$^{-1}$. For this reason, fixing $PA_{\rm kin}$ to a reasonable value is recommended whenever possible to avoid this bias, especially when the output $V_{\rm rot}$ is low.

\begin{figure}
\begin{center}
\includegraphics[width=84mm]{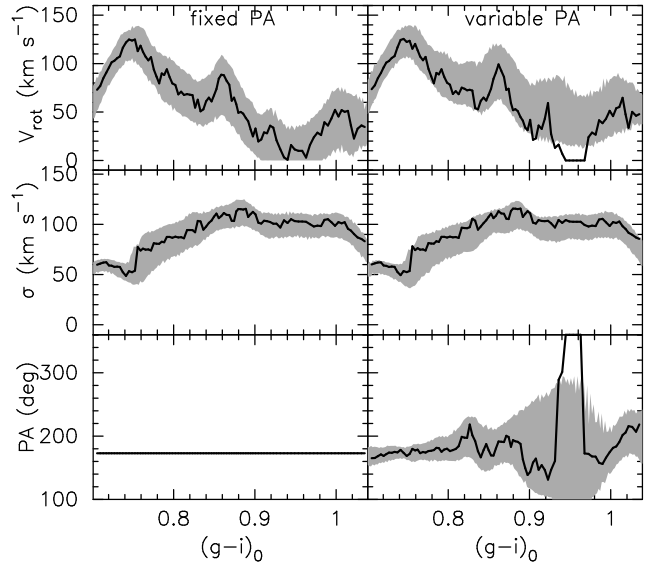}
\caption{Rolling colour fits to the GC kinematics. Two cases are presented: $PA_{\rm kin}=PA_{\rm phot}$ (left panels) and variable $PA_{\rm kin}$ (right panels). Shaded areas represent 68 per cent confidence intervals.}\label{fig:rolcol}
\end{center}
\end{figure}

We find that the kinematic position angle for the blue GCs (170$^{+16}_{-14}$ deg) is consistent with the photometric position angle in the inner parts of the galaxy measured by \citetalias{Napolitano09} (i.e., $PA_{\rm phot}\sim178$ deg, see Fig. \ref{fig:kinemetry}). As will be discussed in great detail below, the red GCs show lower rotation, making the determination of the kinematic PA highly uncertain and artificially enhancing $V_{\rm rot}$. Therefore, we fix the position angle to the photometric position angle (i.e., $PA_{{\rm kin},j}=PA_{\rm phot}$) for both subpopulations in order to better constrain the other parameters. The resulting fits for our fiducial colour split are shown in Fig. \ref{fig:GCkin} for the whole sample and rolling fits as a function of radius can be found in Fig. \ref{fig:kinemetry} for both GC subpopulations. We exclude GC88 from the radial rolling fits as its position at $>8r_e$ skews the measured rolling radius significantly. This does not change the amplitude of the fitted parameters significantly. We find significant major-axis rotation at all radii for the metal-poor GCs only with $41\lesssim V_{rot}\lesssim95$km s$^{-1}$ (see Fig. \ref{fig:kinemetry}). The metal-rich subpopulation may not show significant rotation as $V_{rot}$ is consistent with 0 km s$^{-1}$ at most radii. To test the effect of our colour split assumption, we perform rolling fits as a function of $(g-i)_0$ colour with moving bins of size 25, which we show in Fig. \ref{fig:rolcol}. These demonstrate that intermediate colour GCs (i.e., $0.90<(g-i)_0<0.98$), in particular, do not show significant rotation and that this results in a higher uncertainty on the kinematic position angle. This peculiarity is yet another hint of a possible third intermediate colour GC subpopulation around NGC~4494. On the other hand, the reddest GCs ($(g-i)_0>0.98$) tentatively show some rotation. In every case we find that the velocity dispersion ($\sigma$) of all subpopulations are consistent with each other (see Fig. \ref{fig:kinemetry}).

		\subsubsection{GC metallicities}
We are able to measure spectroscopic metallicities for 54 individual GCs and the three UCDs around NGC~4494 using the CaT and the method described in \citetalias{Foster10}. Briefly, we fit the GC spectra using the {\sc pPXF} routine described in \citet{Cappellari04} to obtain a template of our GC spectra. We use the set of templates described in Section \ref{sec:GCkin} (example fits are shown in Fig. \ref{fig:GCspec}). Next, the fitted spectra are continuum normalised using the {\sc iraf} routine {\sc continuum} with a spline3 function of order $\sim4$ and a stricter lower sigma clipping to avoid spectral features being fitted. The CaT index defined in \citetalias{Foster10} (CaT$_{\rm F10}$, see Table \ref{table:CaT}) is measured on the template fitted and continuum normalised spectra. We emphasise that this ${\rm CaT_{F10}}$ is different from $CaT_{\rm F09}$ used above for galaxy spectra, which has broader passbands to accommodate the broadening of the CaT features due to large velocity dispersion in galaxies. The $CaT_{\rm F10}$ indices are transformed into metallicity using the following equation:
\begin{equation}
{\rm [Fe/H]}=-3.641+0.438\times {\rm CaT}_{\rm F10}
\end{equation}
based on the empirical conversion of \citet{AZ88} derived from the integrated light spectra of Galactic GCs. For more information on this technique, see \citetalias{Foster10}.

\begin{figure}
\begin{center}
\includegraphics[width=84mm]{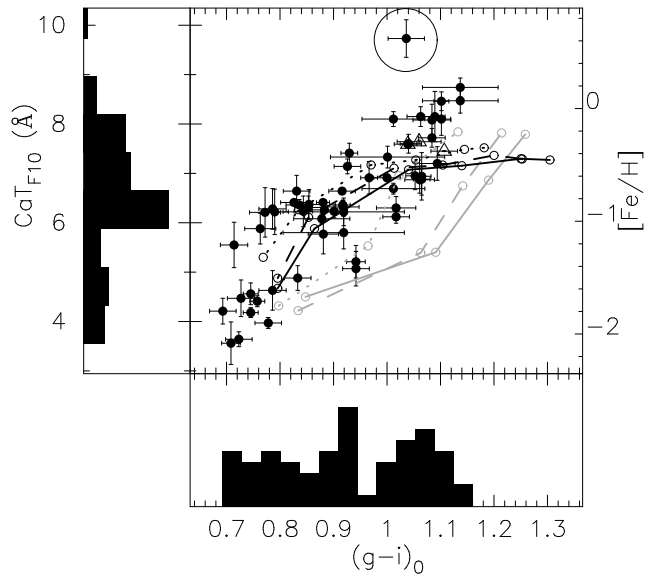} 
\caption{${\rm CaT_{F10}}$ and inferred metallicity (right $y$-axis) as a function of colour for GCs (filled circles) and UCDs (hollow triangles) around NGC~4494. All repeat ${\rm CaT_{F10}}$ measurements of individual objects are shown. Histograms for each axis are also shown. One outlier (highlighted with a large hollow circle) is discussed in the text. Predictions from the single stellar population models of \citetalias{V03} (black) and \citet[][grey]{BC03} are shown for 5 (dotted line), 9 (dashed line) and 13 (solid line) Gyr.}\label{fig:giCaT}
\end{center}
\end{figure}

Fig. \ref{fig:giCaT} shows our measured ${\rm CaT_{F10}}$ index as a function of $(g-i)_0$ colour. We keep multiple ${\rm CaT_{F10}}$ measurements of individual objects as separate data points. Predictions from the \citetalias{V03} and \citet{BC03} simple stellar population models are overlaid for comparison.

There are several immediately striking features in Fig. \ref{fig:giCaT}. First of all, the correlation between the ${\rm CaT_{F10}}$ and colour is obvious and consistent with linear, albeit with large observational scatter. There is one outlier, namely the faint ($i_0=21.8$) red GC, GC102. There is nothing obviously wrong with the photometry or spectrum (fitted or raw) of GC102, so we cannot explain the position of this (low signal-to-noise) outlier in Fig. \ref{fig:giCaT}.

The bulk of the GC data lie above the \citet{BC03} single stellar population models and close to those of \citetalias{V03} (for bluer colours). The apparently linear relationship found in NGC~4494 data up to the reddest colours are in contrast to the findings of \citetalias{Foster10} for the giant elliptical galaxy NGC~1407, where the generally redder data follow the 13 Gyr \citetalias{V03} model track more closely at all colours. Many blue/red GCs in NGC~1407 have higher/lower measured ${\rm CaT_{F10}}$ indices than the NGC~4494 GCs. We discuss this discrepancy and its implications for using the CaT as a metallicity indicator for extragalactic GCs in Appendix \ref{sec:CaTGCs}.

Finally, there appears to be a concentration of GCs with ${\rm CaT_{F10}}\sim6.5$\AA\space (or [Fe/H]$\sim-0.9$). While our sample of spectroscopic GC metallicities is rather small for inferring the global properties of the distribution of GC metallicities in NGC~4494, it appears to be single peaked despite exhibiting clear $(g-i)_0$ colour bimodality (KMM yields a p-value of 0.002 for the confirmed GC sample). We note however that KMM is less reliable for small samples \citep[see][]{Ashman94}. Nevertheless, it is puzzling that the peak in the CaT distribution corresponds to the trough of the colour distribution (see Fig. \ref{fig:giCaT}). This result is reminiscent of that obtained for NGC~1407 (\citetalias{Foster10}) and \citet{Caldwell11}, where the clear colour bimodality also translated into a skewed single-peaked spectroscopic metallicity distribution.

		\subsubsection{Three UCDs around NGC~4494}\label{sec:UCD}

As briefly mentioned in Section \ref{sec:selection}, we report the discovery of 3 spectroscopically confirmed UCDs \citep{Drinkwater00} associated with NGC~4494. We adopt a magnitude definition for UCDs of $M_i<-10.9$, roughly equivalent to that adopted by \citet{Evstigneeva08} in the $V$-band. The 3 UCDs have absolute $i$-band magnitudes of -11.88, -11.92 and -11.71 mag, well within the range for UCD luminosities. Most UCDs have typically been found in dense cluster environments although, for example, one UCD has been confirmed around the Sombrero galaxy \citep{Spitler06,Hau09}, a spiral galaxy in a low-density environment. Another relevant example is that of NGC~ 5128, also an $L^*$ early-type galaxy, wherein several possible UCDs may have been found \citep{Taylor10}.

There is ongoing discussion in the literature about the origin and definition of UCDs. Popular formation scenarios propose that UCDs are either (1) the bright end of the compact star cluster (i.e., GC) luminosity function \citep[e.g.,][]{Mieske04}, or (2) the slightly more extended remnants of tidally stripped dwarf galaxies \citep{Bekki03}. Of course, both scenarios may occur \citep[see][]{DaRocha11,Norris11}. We thus examine the properties of the UCDs around NGC~4494 in order to determine their most likely origin.

UCD1 and UCD2 are within the HST imaging. We measure their sizes using {\sc ishape} \citep{Larsen99} and obtain $1.7\pm0.4$ and $2.3\pm0.6$pc for UCD1 and UCD2, respectively. We are unable to obtain a size estimate for UCD3 as it is unresolved on the Subaru image. Therefore, at least two of the three UCDs around NGC~4494 are compact, as predicted by the star cluster origin scenario. All three UCDs have high CaT-inferred metallicities ($-0.3\gtrsim[Fe/H]\gtrsim-0.4$) as is also the case for some other UCDs \citep[e.g.,][]{Evstigneeva07}. UCD1 is detected with Chandra in the X-ray \citep[][object id NGC\_4494\_CXOU\_12:3125.5+254619]{Humphrey08} with $L_X=(2.1\pm1.1)\times10^{38}$ ergs s$^{-1}$. This X-ray luminosity is consistent with the presence of low mass X-ray binary stars in UCD1.

Following \citet{DaRocha11}, we compute the number of expected GCs brighter than $i_0=19.3$ (i.e., the magnitude of our faintest UCD candidate) around NGC~4494 assuming that UCDs around NGC~4494 are simply the bright end of its GC system. We use the total number of GCs and the GC luminosity function \citep{Kundu01} to estimate that $2\pm2$ UCDs are consistent with being the bright extension of the GC system. Thus, under these assumptions, our 3 confirmed UCDs brighter than $i_0=19.3$ around NGC~4494 are consistent with representing the bright end of the GC luminosity function. We conclude that there is no need to invoke another formation channel such as tidal stripping of dwarf galaxies \citep[e.g.,][]{DaRocha11,Norris10} for the UCDs around NGC~4494.

\section{Discussion}\label{sec:discussion}

%Recent theoretical studies of the formation and evolution of galaxies suggest that a variety of processes are important in shaping and determining the global properties of galaxies. Processes generally invoked in order to reproduce observational constraints within popular theoretical paradigms include galaxy mergers, dissipation of gas, and feedback from either stellar winds, SN or AGN \citep[e.g.,][]{Croton06,Naab07,Hopkins09,Hoffman09,Hoffman10}. Galaxy formation models make predictions for the effects of these processes on the remnant galaxy's stellar population distribution and kinematics but GC system modelling is less advanced. 
In this section, we compare the observed properties of NGC~4494 with the predictions from theoretical models in order to get an understanding of its formation.

\subsection{Inferences from the stellar light}\label{sec:Discussion1}

In Section \ref{sec:stelkin} we report a `flattening' of the stellar kinematics of NGC~4494 with radius such that the kinematics become more disky at large radii. In other words, at large radii only the stars close to the semi-major axis show rotation such that the kinematic axis ratio $q_{\rm kin}$ is low. This is supported by the observed anti-correlation between $h_3$ and $(V_{\rm obs}-V_{\rm sys})/\sigma$, which indicates the presence of a disk-like structure. However, it contrasts with the stellar surface brightness of the galaxy, which has a very constant and relatively round profile at all radii. This kinematic flattening at large radii may be related to the transition suggested by \citet{Hoffman10} between 1-3 $r_e$, where the kinematic signature of the progenitors' disk stars survived. However, it is a puzzle as to how such a `kinematic' disk at large radii could be invisible in the imaging data. Indeed, the axis ratio of kinematic sub-structures are usually found to agree with that of the stellar distribution \citep[e.g.,][]{Krajnovic08}.

The stellar populations don't show any hint of a flattened distribution either. \citet{Denicolo05} reported a central age of 6.7 Gyr with a central metallicity of $[Fe/H]\approx+0.03$ dex for NGC~4494 and \citetalias{Foster09} find no evidence for radial metallicity variations with metallicities roughly constant around $[Fe/H]\gtrsim-0.5$ dex between $\sim 0.2$ and $1.4r_e$. In this work, we find no azimuthal variations in the measured CaT index values (see Fig. \ref{fig:gradient}) or colour. The CaT gradient suggests that the luminosity weighted metallicity of the NGC~4494 stars is either higher than $\sim -0.5$ dex from $0.2r_e$ all the way out to at least $\sim1.6$ $r_e$ and/or that the metallicity gradient is undetectably shallow for the CaT method. The colour gradient in the same radial range suggests a moderate metallicity gradient of $\sim-0.17\pm0.02$ dex per dex.

%We look at the departure of NGC~4494 from the fundamental plane of ETG \citep[e.g.,][]{Djorgovski87,Dressler87} using the method of \citet{Forbes98}. We compute $R=2\log\sigma_0+0.286M_B+0.2\mu_e-3.101$. Values for the central velocity dispersion ($\sigma_0$), the total $B$-band magnitude ($M_B$) and the $B$-band the surface brightness at $r_e$ ($\mu_e$) are taken from the Hyperleda database\footnote{http://leda.univ-lyon1.fr}. We obtain $R\sim-0.37$, indicating that NGC~4494 falls below the fundamental plane of ETGs. Such a low value for $R$ is usually observed for morphologically disturbed galaxies such as obvious merger remnants and is usually associated with young central ages ($\sim1.5$ Gyr). While such a young central age is unlikely based on both the red central colour and the spectroscopic age reported by \citet{Denicolo05} for NGC~4494, it may be an indication that it has gone through a recent interaction.

Recently, the SAURON team reported the results of their 2D stellar population analysis on 48 early-type galaxies \citep{Kuntschner10}. They find that flattened structures in the images of fast-rotators \citep{Emsellem07,Krajnovic08} with disky kinematics have distinct stellar populations, while galaxies classified as slow rotators, and sometimes \emph{harboring inner kinematically decoupled cores}, show no clearly distinct stellar population variation. NGC~4494 does not appear to fit either of those categories. At small radii $\sim 0.1r_e$ , the kinematically decoupled core suggests it is a slow-rotator. At all other radii, its increasingly flattened kinematics suggest it is a fast rotator, but as stated above we find no evidence for a flattening of the stellar distribution or of distinct stellar populations from either the $CaT_{\rm F09}$ or the modelling of the surface brightness profile and colours. In any case, transitions between slow and fast rotators at large radii may be common as \citetalias{Proctor09} also report a transition from a fast to a slow rotator beyond the SAURON field-of-view in NGC~821. Again, these may be the first observational evidences for the transitional kinematics expected to occur between 1 and 3$r_e$ in major merger remnants \citep{Hoffman10}.

\begin{figure}
\begin{center}
\includegraphics[width=84mm]{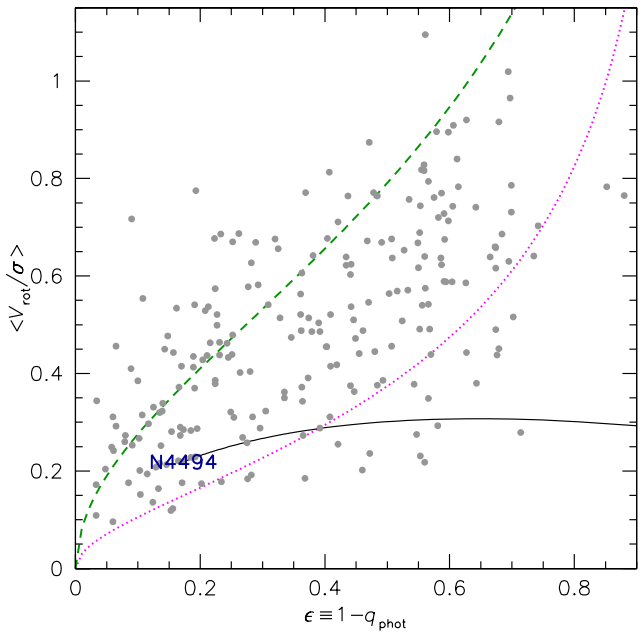} 
\caption{Azimuthally averaged rotation $\left<V_{\rm rot}/\sigma\right>$ versus ellipticity ($\epsilon=1-q_{\rm phot}$) diagnostic diagram, after \citet{Cappellari07}. Data points represent the central regions of nearby early-type galaxies classified  as ``fast-rotators' from the SAURON and ATLAS3D surveys.  The green dashed curve shows a theoretical prediction for edge-on oblate isotropic rotators, and the magenta dotted curve shows the inferred edge-on average relation for nearby fast-rotators.  The observed position of NGC~4494 in the diagram is labelled, with a solid black curve showing the path of possible intrinsic values for a sequence of different assumed inclination angles. This figure is available in colour in the online version.}\label{fig:vsig}
\end{center}
\end{figure}

In Fig. \ref{fig:vsig}, we plot a standard $\left<V_{\rm rot}/\sigma\right>$ vs $\epsilon=1-q_{\rm phot}$ diagnostic diagram \citep{Cappellari07}. This plot can be used to diagnose both the intrinsic structure of the galaxy (in particular NGC~4494) and the nature of the merger.  We have highlighted NGC~4494 on this diagram showing the ATLAS3D data \citep{Emsellem11}. The green curve in Fig.~\ref{fig:vsig} represents an oblate isotropic rotator seen edge-on, and the magenta curve shows what is typical for an edge-on fast rotator after modelling the dynamics of the SAURON sample as inferred by \citet{Cappellari07}. Because galaxies are generally observed \emph{below} the green curve, this suggests that they are \emph{not} isotropic. These results allow us to derive a best-guess solution for the inclination of any fast rotator (including NGC~4494) by assuming that its internal anisotropy follows the mean trend of the other galaxies.  We can also thereby estimate the ellipticity and $\left<V_{\rm rot}/\sigma\right>$ values that would be obtained if the galaxy was viewed {\it edge-on}.

The black curve shows the track of possible edge-on values for NGC~4494 for a series of different inclinations, where we note that the dependence of the dispersion on inclination enters through the assumed anisotropy. The intersection of the black curve with the magenta curve then represents the self-consistent solution for NGC~4494 under the SAURON-based anisotropy assumption. From this we conclude that NGC~4494 is {\it most likely} (but not definitively) a flattened galaxy ($q_{\rm phot}\sim0.6$) seen at an inclination of $\sim45$ degrees. It may even be an S0 rather than a bona fide elliptical. The uncertainties here are driven not by the measured parameters (which are determined very precisely) but by the intrinsic scatter in the anisotropy-ellipticity relation. To estimate this, we use the spread of observed galaxies to the right of the magenta curve in Fig. \ref{fig:vsig}. This suggests an intrinsic ellipticity uncertainty of $\sim\pm0.2$, and an inclination between 39 and 90 degrees. Similarly, using 2D dynamical modelling of NGC~4494, \citet{Rodionov11} found that a $\sim45$ degrees inclination may be preferred. If this interpretation is correct, then to recover edge-on $V_{\rm rot}$ estimates, all of our velocity estimates should be increased by $\sim$~40 per cent.

\subsection{Inferences from the GC system}
We find that the GC colour subpopulations are reasonably well defined in NGC~4494 and choose a nominal colour cut at $(g-i)_0=0.99$ to delineate metal-poor from -rich GCs. The 54 spectroscopically measured GC metallicities vary between -2.0 dex $\lesssim[Fe/H]\lesssim0.0$ dex and their distribution appears single-peaked around $[Fe/H]_{mean}\sim -1.0$ dex despite their clear bimodal colour distribution. A similar behaviour was found for GCs around NGC~1407 (\citetalias{Foster10}) and M31 \citep{Caldwell11}. These may have some interesting implications for the ubiquity of the GC metallicity bimodality as inferred from GC colour distributions \citep[also see][]{Yoon06,Peng06,Blakeslee10}.

It has been suggested (and shown) that the distribution of red GCs usually follows that of the galaxy stars, while blue GCs follow the X-ray halo profile of their host galaxy \citep[e.g.,][]{Minniti96,Forbes04,Boley09}. This suggestion is based on both model predictions and the similarities of the respective spatial distribution and abundances. Surprisingly, we find that the surface brightness profile of the stars does not compare well with the spatial distribution of the red (or the blue) GCs in NGC~4494 (see Fig. \ref{fig:GCdens}). Indeed, the Sers\'ic index of the stars is inconsistent with that of both GC subpopulations, and this difference is even more pronounced for the red GCs. A similar conclusion is reached from comparing the kinematics of the stars, which agrees better with the blue GC kinematics than with that of the red GCs at the same radius (see Fig. \ref{fig:kinemetry}). On the other hand, the colour of the galaxy stars is consistent with that of the red GCs (see Fig. \ref{fig:gradient}). Therefore, the association of the red GCs with the galaxy stars is less clear in NGC~4494 than previously observed in other galaxies.

\begin{figure}
\begin{center}
\includegraphics[width=84mm]{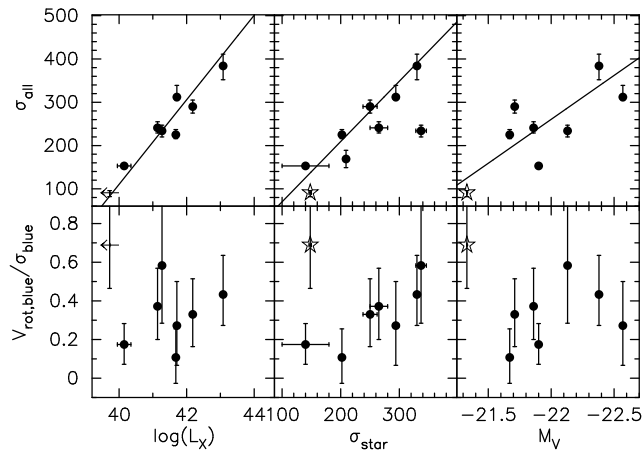}
\caption{Comparison of salient GC kinematic properties with host galaxy global parameters: X-ray luminosity  ($\log({L_X})$), stellar velocity dispersion ($\sigma_{star}$) and absolute $V$-band magnitude ($M_V$). Symbols on the y-axis and GC kinematics are from Table \ref{table:litGCkin}. We include literature GC kinematic studies (filled circles) and NGC~4494 data (hollow stars and upper limit symbols). Solid lines are bisector fits to correlated parameters from \citet[][figure 12]{Lee10}. Datum for NGC~4494 disagrees with the $V_{\rm rot,blue}/\sigma_{\rm blue}$ vs $\sigma_{star}$ trend observed in other large elliptical galaxies (middle lower panel).}\label{fig:lee}
\end{center}
\end{figure}

In Section \ref{sec:GCkin}, we measure the rotational velocity and velocity dispersion of the blue and red GCs. In order to put our results into a broader context, Fig. \ref{fig:lee} reproduces parts of figure 12 from \citet{Lee10} using the updated kinematic data presented in Table \ref{table:litGCkin} and compares GC kinematics with the global properties of the host galaxies for large ETGs. One caveat of the following comparison has to do with the heterogeneity of the methods used in the various dynamical studies represented in Fig. \ref{fig:lee}. Moreover, NGC~4494 is the only fast-rotator with large GC kinematic sample. Thus, its GC kinematics may be intrinsically different. Nevertheless, we find that NGC~4494 compares well with other large ETGs. We confirm that it agrees with the trends found between the velocity dispersion of the whole GC system and X-ray, central galactic velocity dispersion and $B$-band absolute magnitude as reported in \citet{Lee10}. However, the kinematics of the blue GCs of NGC~4494 appear to deviate from that of other large elliptical galaxies in the $V_{\rm rot,blue}/\sigma_{\rm blue}$ vs $\sigma_{star}$ space, emphasising the large rotation of the blue GCs. Perhaps surprisingly given the arguably ``unusually low'' X-ray luminosity of NGC~4494, the GC kinematics agree well with the expected trend with galaxy X-ray luminosity \citep[e.g.,][]{Romanowsky03,OSullivan04}. This also suggests that the processes involved in the formation of NGC~4494 that led to its peculiar X-ray luminosity and kinematics have preserved the majority of the correlations between its global properties and its GC kinematics as with other large ETGs.

Assuming that NGC~4494 has indeed undergone a recent interaction as inferred by \citet{OSullivan04} from its X-ray luminosity, then we may reasonably conclude that the bulk of its GC system formed before that interaction and is thus a conglomerate of the progenitors' GC systems. Therefore, the current GC kinematics may hold clues to the understanding of the kinematics of the progenitor galaxies involved in the latest merger event. In this spirit, we use numerical simulations of disk-disk major mergers in order to see what can be learned about the progenitors from the GC kinematics. Details of the simulations and results are presented in Appendix \ref{sec:Kenji}. By studying numerous simulations of mergers with a variety of initial conditions, we find that the model that best reproduces the observed kinematics of the blue and red GCs (ignoring the intermediate colour GCs that show little rotation) involves a major disk-disk merger with a large amount of orbital angular momentum and a retrograde-retrograde orbital configuration. Models with other orbital configurations do not yield GC kinematics that are consistent with what is observed for NGC~4494. This is a suggestion that NGC~4494 may be a major merger remnant.

\section{Summary and conclusions}\label{sec:conclusion}
NGC~4494 has often been dubbed an `ordinary elliptical' galaxy, making it an ideal target for understanding the formation and evolution of a typical galaxy. In this work, we combine imaging and spectroscopy of stellar light and GCs in NGC~4494. From the imaging we obtain the stellar surface brightness profiles and GC density profiles. We find that the colour distribution of the GC system around NGC~4494 is statistically better fitted by two Gaussians than a single one suggesting metallicity bimodality. The spectroscopy yields spatially resolved kinematics and abundances of the stellar component and GCs. A total of 109 individual GCs and 3 UCD/DGTOs are confirmed spectroscopically. The properties of the UCDs are consistent with them being bright GCs. Metallicities are measured for 54 GCs and the 3 UCDs using the CaT absorption lines. The CaT inferred metallicity distribution for the GC system is single-peaked, in contrast to that inferred from the colours, which is clearly bimodal. Measurements for both the stellar light and GC spectroscopy can be found in Appendix \ref{sec:datatables}. We find that while the intermediate colour (green) GCs do not rotate, the blue (and possibly reddest) GCs do rotate (Fig. \ref{fig:rolcol}) as do the galaxy stars (Fig. \ref{fig:kinemetry}). The velocity dispersion of all GC subpopulations and the galaxy stars are consistent.

A comparison of the GC system's kinematics and the global properties of NGC~4494 with those of other large ETGs suggest that NGC~4494 is indeed typical with the exception of the unusually high blue GC rotation. We find suggestive evidence in the distinct kinematics for a possible third intermediate colour (green) GC subpopulation. We infer that most of the observational evidence suggests that NGC~4494 is consistent with formation via a recent \emph{gas-rich} major merger, but other formation scenarios cannot be ruled out. Some remaining open issues are:
\begin{enumerate}
\item the relatively old (6.7 Gyr) measured central age suggests that little dissipation and recent central star formation has occurred (i.e., possibly a dry merger). However, its cuspy surface brightness profile and the presence of a kinematically decoupled core suggest a high gas fraction of $\gtrsim15$ per cent \citep{Hopkins09,Hoffman10}. This is also supported by the presence of an inner dust ring \citep{Xu10}. Therefore, the evidence is not conclusive as to the amount of gas dissipation involved in the formation of NGC~4494. The possibly `unusually' low X-ray luminosity suggests that the mechanism involved in depleting the gas must have allowed enough dissipation to form a kinematically decoupled core without significant recent central star formation.
\item The inner kinematic profile of NGC~4494 follows the light profile well up to $\sim1.8r_e$, beyond that the kinematics become increasingly flattened with radius (i.e., the axis ratio $q_{\rm kin}$ is low). We find no evidence of a flattened component in the stellar distribution from our imaging. It is however not surprising to find transitions in the kinematics of merger remnants at 1-3$r_e$ as they are predicted in the models of \citet{Hoffman10}.  We suggest that NGC~4494 may be a flattened galaxy ($q_{\rm phot}=0.6\pm0.2$), possibly even an S0, seen at an inclination of $45^{+45}_{-6}$ degrees based on its position in the classic $\left<V_{\rm rot}/\sigma\right>$ vs $\epsilon=1-q_{\rm phot}$ diagram (Fig \ref{fig:vsig}). This may explain its high flattened rotation with round photometric isophotes.
\end{enumerate}

Off-axis deep optical spectroscopy out to large galactocentric radii would be a good way to independently test the flattened kinematics. It would also allow for a stellar population analysis (i.e., ages, metallicities and ${\rm\alpha}$-element abundances) using the standard and well tested Lick system \citep[e.g.,][]{Proctor02}. The added knowledge of spatially resolved stellar ages and ${\rm\alpha}$-element abundances may highlight distinct stellar populations as observed by SAURON \citep{Kuntschner10} that were not seen from our near-infrared spectra.

This work shows that such complete studies of individual galaxies incorporating as many lines of evidence as available can help disentangle the processes involved in the formation and evolution of selected galaxies. As such datasets become available, observational constraints based on statistical samples will help improve our understanding of the details of galaxy formation and evolution.

\section*{Acknowledgements}
We would like to thank the anonymous Referee for his/her careful reading of our manuscript and helpful comments. Soeren Larsen is acknowledged for providing us with the HST/WFPC2 GC catalogue. We thank Michele Cappellari and Loren Hoffman for helpful discussions. CF thanks the Anglo-Australian Observatory for financial support in the form of a graduate top-up scholarship. DAF thanks the ARC for financial support. JPB and AJR acknowledge support from NSF grant AST-0808099 and AST-0909237. Based in part on data collected at Subaru Telescope and obtained from the SMOKA, which is operated by the Astronomy Data Center, National Astronomical Observatory of Japan. Part of the Subaru data analysed was obtained through a Gemini time exchange, program GN-2008A-C-12. The data presented herein were obtained at the W.M. Keck Observatory, which is operated as a scientific partnership among the California Institute of Technology, the University of California and the National Aeronautics and Space Administration. The Observatory was made possible by the generous financial support of the W.M. Keck Foundation. The analysis pipeline used to reduce the DEIMOS data was developed at UC Berkeley with support from NSF grant AST-0071048. We acknowledge the usage of the HyperLeda database (http://leda.univ-lyon1.fr) and of NASA/IPAC Extragalactic Database (NED), which is operated by the Jet Propulsion Laboratory, California Institute of Technology, under contract with the National Aeronautics and Space Administration.

\bibliographystyle{mn2e}
\bibliography{biblio}

\begin{thebibliography}{}

\bibitem[\protect\citeauthoryear{{Abazajian}, {Adelman-McCarthy},
  {Ag{\"u}eros}, {Allam}, {Allende Prieto}, {An}, {Anderson}, {Anderson},
  {Annis}, {Bahcall} \& et al.}{{Abazajian} et~al.}{2009}]{Abazajian09}
{Abazajian} K.~N.,   {Adelman-McCarthy} J.~K., et al. 2009, \apjs, 182, 543

\bibitem[\protect\citeauthoryear{{Armandroff} \& {Zinn}}{{Armandroff} \&
  {Zinn}}{1988}]{AZ88}
{Armandroff} T.~E.,  {Zinn} R.,  1988, \aj, 96, 92

\bibitem[\protect\citeauthoryear{{Ashman}, {Bird} \& {Zepf}}{{Ashman}
  et~al.}{1994}]{Ashman94}
{Ashman} K.~M.,  {Bird} C.~M.,    {Zepf} S.~E.,  1994, \aj, 108, 2348

\bibitem[\protect\citeauthoryear{{Ashman} \& {Zepf}}{{Ashman} \&
  {Zepf}}{1992}]{Ashman92}
{Ashman} K.~M.,  {Zepf} S.~E.,  1992, \apj, 384, 50

\bibitem[\protect\citeauthoryear{{Bassino}, {Richtler} \& {Dirsch}}{{Bassino}
  et~al.}{2006}]{Bassino06}
{Bassino} L.~P.,  {Richtler} T.,    {Dirsch} B.,  2006, \mnras, 367, 156

\bibitem[\protect\citeauthoryear{{Beasley}, {Baugh}, {Forbes}, {Sharples} \&
  {Frenk}}{{Beasley} et~al.}{2002}]{Beasley02}
{Beasley} M.~A.,  {Baugh} C.~M.,  {Forbes} D.~A.,  {Sharples} R.~M.,    {Frenk}
  C.~S.,  2002, \mnras, 333, 383

\bibitem[\protect\citeauthoryear{{Bekki}}{{Bekki}}{2010}]{Bekki10}
{Bekki} K.,  2010, \mnras, 401, 2753

\bibitem[\protect\citeauthoryear{{Bekki}, {Beasley}, {Brodie} \&
  {Forbes}}{{Bekki} et~al.}{2005}]{Bekki05}
{Bekki} K.,  {Beasley} M.~A.,  {Brodie} J.~P.,    {Forbes} D.~A.,  2005,
  \mnras, 363, 1211

\bibitem[\protect\citeauthoryear{{Bekki}, {Couch}, {Drinkwater} \&
  {Shioya}}{{Bekki} et~al.}{2003}]{Bekki03}
{Bekki} K.,  {Couch} W.~J.,  {Drinkwater} M.~J.,    {Shioya} Y.,  2003, \mnras,
  344, 399

\bibitem[\protect\citeauthoryear{{Bekki}, {Forbes}, {Beasley} \&
  {Couch}}{{Bekki} et~al.}{2002}]{Bekki02}
{Bekki} K.,  {Forbes} D.~A.,  {Beasley} M.~A.,    {Couch} W.~J.,  2002, \mnras,
  335, 1176

\bibitem[\protect\citeauthoryear{{Bekki} \& {Shioya}}{{Bekki} \&
  {Shioya}}{1999}]{Bekki99}
{Bekki} K.,  {Shioya} Y.,  1999, \apj, 513, 108

\bibitem[\protect\citeauthoryear{{Bell}, {McIntosh}, {Katz} \&
  {Weinberg}}{{Bell} et~al.}{2003}]{Bell03}
{Bell} E.~F.,  {McIntosh} D.~H.,  {Katz} N.,    {Weinberg} M.~D.,  2003, \apjs,
  149, 289

\bibitem[\protect\citeauthoryear{{Bender}, {Saglia} \& {Gerhard}}{{Bender}
  et~al.}{1994}]{Bender94}
{Bender} R.,  {Saglia} R.~P.,    {Gerhard} O.~E.,  1994, \mnras, 269, 785

\bibitem[\protect\citeauthoryear{{Bergond}, {Zepf}, {Romanowsky}, {Sharples} \&
  {Rhode}}{{Bergond} et~al.}{2006}]{Bergond06}
{Bergond} G.,  {Zepf} S.~E.,  {Romanowsky} A.~J.,  {Sharples} R.~M.,    {Rhode}
  K.~L.,  2006, \aap, 448, 155

\bibitem[\protect\citeauthoryear{{Bica} \& {Alloin}}{{Bica} \&
  {Alloin}}{1987}]{Bica87}
{Bica} E.,  {Alloin} D.,  1987, \aap, 186, 49

\bibitem[\protect\citeauthoryear{{Blakeslee}, {Cantiello} \&
  {Peng}}{{Blakeslee} et~al.}{2010}]{Blakeslee10}
{Blakeslee} J.~P.,  {Cantiello} M.,    {Peng} E.~W.,  2010, \apj, 710, 51

\bibitem[\protect\citeauthoryear{{Boley}, {Lake}, {Read} \& {Teyssier}}{{Boley}
  et~al.}{2009}]{Boley09}
{Boley} A.~C.,  {Lake} G.,  {Read} J.,    {Teyssier} R.,  2009, \apjl, 706,
  L192

\bibitem[\protect\citeauthoryear{{Brodie} \& {Strader}}{{Brodie} \&
  {Strader}}{2006}]{Brodie06}
{Brodie} J.~P.,  {Strader} J.,  2006, \araa, 44, 193

\bibitem[\protect\citeauthoryear{{Brodie}, {Strader}, {Denicol{\'o}},
  {Beasley}, {Cenarro}, {Larsen}, {Kuntschner} \& {Forbes}}{{Brodie}
  et~al.}{2005}]{Brodie05}
{Brodie} J.~P.,   {Strader} J., et al. 2005, \aj, 129,   2643

\bibitem[\protect\citeauthoryear{{Bruzual} \& {Charlot}}{{Bruzual} \&
  {Charlot}}{2003}]{BC03}
{Bruzual} G.,  {Charlot} S.,  2003, \mnras, 344, 1000

\bibitem[\protect\citeauthoryear{{Burkert} \& {Tremaine}}{{Burkert} \&
  {Tremaine}}{2010}]{Burkert10}
{Burkert} A.,  {Tremaine} S.,  2010, \apj, 720, 516

\bibitem[\protect\citeauthoryear{{Caldwell}, {Schiavon}, {Morrison}, {Rose} \&
  {Harding}}{{Caldwell} et~al.}{2011}]{Caldwell11}
{Caldwell} N.,  {Schiavon} R.,  {Morrison} H.,  {Rose} J.~A.,    {Harding} P.,
  2011, \aj, 141, 61

\bibitem[\protect\citeauthoryear{{Capaccioli}, {Caon} \&
  {D'Onofrio}}{{Capaccioli} et~al.}{1992}]{Capaccioli92}
{Capaccioli} M.,  {Caon} N.,    {D'Onofrio} M.,  1992, \mnras, 259, 323

\bibitem[\protect\citeauthoryear{{Cappellari} \& {Emsellem}}{{Cappellari} \&
  {Emsellem}}{2004}]{Cappellari04}
{Cappellari} M.,  {Emsellem} E.,  2004, \pasp, 116, 138

\bibitem[\protect\citeauthoryear{{Cappellari}, {Emsellem}, {Bacon}, {Bureau},
  {Davies}, {de Zeeuw}, {Falc{\'o}n-Barroso}, {Krajnovi{\'c}}, {Kuntschner},
  {McDermid}, {Peletier}, {Sarzi} \& et al.}{{Cappellari}
  et~al.}{2007}]{Cappellari07}
{Cappellari} M.,   {Emsellem} E., et al. 2007, \mnras, 379, 418

\bibitem[\protect\citeauthoryear{{Cenarro}, {Beasley}, {Strader}, {Brodie} \&
  {Forbes}}{{Cenarro} et~al.}{2007}]{Cenarro07}
{Cenarro} A.~J.,  {Beasley} M.~A.,  {Strader} J.,  {Brodie} J.~P.,    {Forbes}
  D.~A.,  2007, \aj, 134, 391

\bibitem[\protect\citeauthoryear{{Cenarro}, {Cardiel} \& {Gorgas}}{{Cenarro}
  et~al.}{2008}]{Cenarro08a}
{Cenarro} A.~J.,  {Cardiel} N.,    {Gorgas} J.,  2008, in {J.~H.~Knapen,
  T.~J.~Mahoney, \& A.~Vazdekis} ed., Pathways Through an Eclectic Universe
  Vol.~390 of Astronomical Society of the Pacific Conference Series, {The
  Calcium Triplet Gradient of M32}.
pp 292--+

\bibitem[\protect\citeauthoryear{{Coccato}, {Gerhard}, {Arnaboldi}, {Das},
  {Douglas}, {Kuijken}, {Merrifield}, {Napolitano}, {Noordermeer},
  {Romanowsky}, {Capaccioli}, {Cortesi}, {de Lorenzi} \& {Freeman}}{{Coccato}
  et~al.}{2009}]{Coccato09}
{Coccato} L.,   {Gerhard} O., et al. 2009, \mnras, 394, 1249

\bibitem[\protect\citeauthoryear{{C{\^o}t{\'e}}, {McLaughlin}, {Cohen} \&
  {Blakeslee}}{{C{\^o}t{\'e}} et~al.}{2003}]{Cote03}
{C{\^o}t{\'e}} P.,  {McLaughlin} D.~E.,  {Cohen} J.~G.,    {Blakeslee} J.~P.,
  2003, \apj, 591, 850

\bibitem[\protect\citeauthoryear{{C{\^o}t{\'e}}, {McLaughlin}, {Hanes},
  {Bridges}, {Geisler}, {Merritt}, {Hesser}, {Harris} \& {Lee}}{{C{\^o}t{\'e}}
  et~al.}{2001}]{Cote01}
{C{\^o}t{\'e}} P.,   {McLaughlin} D.~E., et al. 2001, \apj, 559, 828

\bibitem[\protect\citeauthoryear{{Da Rocha}, {Mieske}, {Georgiev}, {Hilker},
  {Ziegler} \& {Mendes de Oliveira}}{{Da Rocha} et~al.}{2011}]{DaRocha11}
{Da Rocha} C.,   {Mieske} S., et al. 2011, \aap, 525, A86+

\bibitem[\protect\citeauthoryear{{de Vaucouleurs}}{{de
  Vaucouleurs}}{1953}]{deVaucouleurs53}
{de Vaucouleurs} G.,  1953, \mnras, 113, 134

\bibitem[\protect\citeauthoryear{{de Vaucouleurs}, {de Vaucouleurs}, {Corwin}
  Jr., {Buta}, {Paturel} \& {Fouque}}{{de Vaucouleurs}
  et~al.}{1991}]{deVaucouleurs91}
{de Vaucouleurs} G.,  {de Vaucouleurs} A.,  {Corwin} Jr. H.~G.,  {Buta} R.~J.,
  {Paturel} G.,    {Fouque} P.,  1991, {Third Reference Catalogue of Bright
  Galaxies}

\bibitem[\protect\citeauthoryear{{Denicol{\'o}}, {Terlevich}, {Terlevich},
  {Forbes} \& {Terlevich}}{{Denicol{\'o}} et~al.}{2005}]{Denicolo05}
{Denicol{\'o}} G.,  {Terlevich} R.,  {Terlevich} E.,  {Forbes} D.~A.,
  {Terlevich} A.,  2005, \mnras, 358, 813

\bibitem[\protect\citeauthoryear{{Di Matteo}, {Pipino}, {Lehnert}, {Combes} \&
  {Semelin}}{{Di Matteo} et~al.}{2009}]{DiMatteo09}
{Di Matteo} P.,  {Pipino} A.,  {Lehnert} M.~D.,  {Combes} F.,    {Semelin} B.,
  2009, \aap, 499, 427

\bibitem[\protect\citeauthoryear{{Diaz}, {Terlevich} \& {Terlevich}}{{Diaz}
  et~al.}{1989}]{Diaz89}
{Diaz} A.~I.,  {Terlevich} E.,    {Terlevich} R.,  1989, \mnras, 239, 325

\bibitem[\protect\citeauthoryear{{Dirsch}, {Richtler}, {Geisler}, {Forte},
  {Bassino} \& {Gieren}}{{Dirsch} et~al.}{2003}]{Dirsch03}
{Dirsch} B.,   {Richtler} T., et al. 2003, \aj, 125, 1908

\bibitem[\protect\citeauthoryear{{Drinkwater}, {Jones}, {Gregg} \&
  {Phillipps}}{{Drinkwater} et~al.}{2000}]{Drinkwater00}
{Drinkwater} M.~J.,  {Jones} J.~B.,  {Gregg} M.~D.,    {Phillipps} S.,  2000,
  \pasa, 17, 227

\bibitem[\protect\citeauthoryear{{Dufour}, {Harvel}, {Martins}, {Schiffer} III,
  {Talent}, {Wells}, {van den Bergh} \& {Talbot} Jr.}{{Dufour}
  et~al.}{1979}]{Dufour79}
{Dufour} R.~J.,   {Harvel} C.~A., et al. 1979, \aj, 84, 284

\bibitem[\protect\citeauthoryear{{Emsellem}, {Cappellari}, {Krajnovi{\'c}},
  {Alatalo}, {Blitz}, {Bois}, {Bournaud}, {Bureau}, {Davies}, {Davis}, {de
  Zeeuw} \& et al.}{{Emsellem} et~al.}{2011}]{Emsellem11}
{Emsellem} E.,   {Cappellari} M., et al. 2011, ArXiv e-prints

\bibitem[\protect\citeauthoryear{{Emsellem}, {Cappellari}, {Krajnovi{\'c}},
  {van de Ven}, {Bacon}, {Bureau}, {Davies}, {de Zeeuw}, {Falc{\'o}n-Barroso},
  {Kuntschner}, {McDermid}, {Peletier} \& {Sarzi}}{{Emsellem}
  et~al.}{2007}]{Emsellem07}
{Emsellem} E.,   {Cappellari} M., et al. 2007, \mnras, 379, 401

\bibitem[\protect\citeauthoryear{{Emsellem}, {Cappellari}, {Peletier},
  {McDermid}, {Bacon}, {Bureau}, {Copin}, {Davies}, {Krajnovi{\'c}},
  {Kuntschner}, {Miller} \& {de Zeeuw}}{{Emsellem} et~al.}{2004}]{Emsellem04}
{Emsellem} E.,   {Cappellari} M., et al. 2004, \mnras, 352,   721

\bibitem[\protect\citeauthoryear{{Evstigneeva}, {Drinkwater}, {Peng}, {Hilker},
  {De Propris}, {Jones}, {Phillipps}, {Gregg} \& {Karick}}{{Evstigneeva}
  et~al.}{2008}]{Evstigneeva08}
{Evstigneeva} E.~A.,   {Drinkwater} M.~J., et al. 2008, \aj, 136, 461

\bibitem[\protect\citeauthoryear{{Evstigneeva}, {Gregg}, {Drinkwater} \&
  {Hilker}}{{Evstigneeva} et~al.}{2007}]{Evstigneeva07}
{Evstigneeva} E.~A.,  {Gregg} M.~D.,  {Drinkwater} M.~J.,    {Hilker} M.,
  2007, \aj, 133, 1722

\bibitem[\protect\citeauthoryear{{Feigelson} \& {Babu}}{{Feigelson} \&
  {Babu}}{1992}]{Feigelson92}
{Feigelson} E.~D.,  {Babu} G.~J.,  1992, \apj, 397, 55

\bibitem[\protect\citeauthoryear{{Forbes}, {Franx}, {Illingworth} \&
  {Carollo}}{{Forbes} et~al.}{1996}]{Forbes96}
{Forbes} D.~A.,  {Franx} M.,  {Illingworth} G.~D.,    {Carollo} C.~M.,  1996,
  \apj, 467, 126

\bibitem[\protect\citeauthoryear{{Forbes}, {Grillmair}, {Williger}, {Elson} \&
  {Brodie}}{{Forbes} et~al.}{1998}]{Forbes98a}
{Forbes} D.~A.,  {Grillmair} C.~J.,  {Williger} G.~M.,  {Elson} R.~A.~W.,
  {Brodie} J.~P.,  1998, \mnras, 293, 325

\bibitem[\protect\citeauthoryear{{Forbes}, {Raul Faifer}, {Carlos Forte},
  {Bridges}, {Beasley}, {Gebhardt}, {Hanes}, {Sharples} \& {Zepf}}{{Forbes}
  et~al.}{2004}]{Forbes04}
{Forbes} D.~A.,   {Raul Faifer} F., et al. 2004, \mnras, 355, 608

\bibitem[\protect\citeauthoryear{{Foster}, {Forbes}, {Proctor}, {Strader},
  {Brodie} \& {Spitler}}{{Foster} et~al.}{2010}]{Foster10}
{Foster} C.,   {Forbes} D.~A., et al. 2010, \aj, 139, 1566

\bibitem[\protect\citeauthoryear{{Foster}, {Proctor}, {Forbes}, {Spolaor},
  {Hopkins} \& {Brodie}}{{Foster} et~al.}{2009}]{Foster09}
{Foster} C.,   {Proctor} R.~N., et al. 2009, \mnras, 400, 2135

\bibitem[\protect\citeauthoryear{{Graham} \& {Driver}}{{Graham} \&
  {Driver}}{2005}]{Graham05}
{Graham} A.~W.,  {Driver} S.~P.,  2005, \pasa, 22, 118

\bibitem[\protect\citeauthoryear{{Harris} \& {Harris}}{{Harris} \&
  {Harris}}{2011}]{Harris11}
{Harris} G.~L.~H.,  {Harris} W.~E.,  2011, \mnras, 410, 2347

\bibitem[\protect\citeauthoryear{{Harris}}{{Harris}}{2009}]{Harris09a}
{Harris} W.~E.,  2009, \apj, 703, 939

\bibitem[\protect\citeauthoryear{{Harris} \& {van den Bergh}}{{Harris} \& {van
  den Bergh}}{1981}]{Harris81}
{Harris} W.~E.,  {van den Bergh} S.,  1981, \aj, 86, 1627

\bibitem[\protect\citeauthoryear{{Hau}, {Spitler}, {Forbes}, {Proctor},
  {Strader}, {Mendel}, {Brodie} \& {Harris}}{{Hau} et~al.}{2009}]{Hau09}
{Hau} G.~K.~T.,   {Spitler} L.~R., et al. 2009, \mnras, 394,   L97

\bibitem[\protect\citeauthoryear{{Hoffman}, {Cox}, {Dutta} \&
  {Hernquist}}{{Hoffman} et~al.}{2010}]{Hoffman10}
{Hoffman} L.,  {Cox} T.~J.,  {Dutta} S.,    {Hernquist} L.,  2010, \apj, 723,
  818

\bibitem[\protect\citeauthoryear{{Hopkins}, {Bundy}, {Croton}, {Hernquist},
  {Keres}, {Khochfar}, {Stewart}, {Wetzel} \& {Younger}}{{Hopkins}
  et~al.}{2010}]{Hopkins10}
{Hopkins} P.~F.,   {Bundy} K., et al. 2010, \apj,   715, 202

\bibitem[\protect\citeauthoryear{{Hopkins}, {Cox}, {Dutta}, {Hernquist},
  {Kormendy} \& {Lauer}}{{Hopkins} et~al.}{2009}]{Hopkins09}
{Hopkins} P.~F.,   {Cox} T.~J., et al. 2009, \apjs, 181, 135

\bibitem[\protect\citeauthoryear{{Humphrey} \& {Buote}}{{Humphrey} \&
  {Buote}}{2008}]{Humphrey08}
{Humphrey} P.~J.,  {Buote} D.~A.,  2008, \apj, 689, 983

\bibitem[\protect\citeauthoryear{{Hwang}, {Lee}, {Park}, {Kim}, {Park}, {Sohn},
  {Lee}, {Rey}, {Lee} \& {Kim}}{{Hwang} et~al.}{2008}]{Hwang08}
{Hwang} H.~S.,   {Lee} M.~G., et al. 2008, \apj, 674, 869

\bibitem[\protect\citeauthoryear{{Jensen}, {Tonry}, {Barris}, {Thompson},
  {Liu}, {Rieke}, {Ajhar} \& {Blakeslee}}{{Jensen} et~al.}{2003}]{Jensen03}
{Jensen} J.~B.,   {Tonry} J.~L., et al. 2003, \apj,   583, 712

\bibitem[\protect\citeauthoryear{{Jorgensen}, {Carlsson} \&
  {Johnson}}{{Jorgensen} et~al.}{1992}]{Jorgensen92}
{Jorgensen} U.~G.,  {Carlsson} M.,    {Johnson} H.~R.,  1992, \aap, 254, 258

\bibitem[\protect\citeauthoryear{{Kissler-Patig} \& {Gebhardt}}{{Kissler-Patig}
  \& {Gebhardt}}{1998}]{Kissler-Patig98}
{Kissler-Patig} M.,  {Gebhardt} K.,  1998, \aj, 116, 2237

\bibitem[\protect\citeauthoryear{{Kobayashi} \& {Arimoto}}{{Kobayashi} \&
  {Arimoto}}{1999}]{Kobayashi99}
{Kobayashi} C.,  {Arimoto} N.,  1999, \apj, 527, 573

\bibitem[\protect\citeauthoryear{{Koch}, {Grebel}, {Wyse}, {Kleyna},
  {Wilkinson}, {Harbeck}, {Gilmore} \& {Evans}}{{Koch} et~al.}{2006}]{Koch06}
{Koch} A.,   {Grebel} E.~K., et al. 2006, \aj, 131,   895

\bibitem[\protect\citeauthoryear{{Krajnovi{\'c}}, {Bacon}, {Cappellari},
  {Davies}, {de Zeeuw}, {Emsellem}, {Falc{\'o}n-Barroso}, {Kuntschner},
  {McDermid}, {Peletier}, {Sarzi}, {van den Bosch} \& {van de
  Ven}}{{Krajnovi{\'c}} et~al.}{2008}]{Krajnovic08}
{Krajnovi{\'c}} D.,   {Bacon} R., et al. 2008, \mnras, 390, 93

\bibitem[\protect\citeauthoryear{{Krajnovi{\'c}}, {Cappellari}, {de Zeeuw} \&
  {Copin}}{{Krajnovi{\'c}} et~al.}{2006}]{Krajnovic06}
{Krajnovi{\'c}} D.,  {Cappellari} M.,  {de Zeeuw} P.~T.,    {Copin} Y.,  2006,
  \mnras, 366, 787

\bibitem[\protect\citeauthoryear{{Krajnovic}, {Emsellem}, {Cappellari},
  {Alatalo}, {Blitz}, {Bois}, {Bournaud}, {Bureau}, {Davies}, {Davis}, {de
  Zeeuw} \& et al.}{{Krajnovic} et~al.}{2011}]{Krajnovic11}
{Krajnovic} D.,   {Emsellem} E., et al. 2011, ArXiv e-prints

\bibitem[\protect\citeauthoryear{{Kundu} \& {Whitmore}}{{Kundu} \&
  {Whitmore}}{2001}]{Kundu01}
{Kundu} A.,  {Whitmore} B.~C.,  2001, \aj, 121, 2950

\bibitem[\protect\citeauthoryear{{Kuntschner}, {Emsellem}, {Bacon},
  {Cappellari}, {Davies}, {de Zeeuw}, {Falc{\'o}n-Barroso}, {Krajnovi{\'c}},
  {McDermid}, {Peletier}, {Sarzi}, {Shapiro}, {van den Bosch} \& {van de
  Ven}}{{Kuntschner} et~al.}{2010}]{Kuntschner10}
{Kuntschner} H.,   {Emsellem} E., et al. 2010, \mnras, 408, 97

\bibitem[\protect\citeauthoryear{{Lackner} \& {Ostriker}}{{Lackner} \&
  {Ostriker}}{2010}]{Lackner10}
{Lackner} C.~N.,  {Ostriker} J.~P.,  2010, \apj, 712, 88

\bibitem[\protect\citeauthoryear{{Larsen}}{{Larsen}}{1999}]{Larsen99}
{Larsen} S.~S.,  1999, \aaps, 139, 393

\bibitem[\protect\citeauthoryear{{Larsen}, {Brodie}, {Huchra}, {Forbes} \&
  {Grillmair}}{{Larsen} et~al.}{2001}]{Larsen01}
{Larsen} S.~S.,  {Brodie} J.~P.,  {Huchra} J.~P.,  {Forbes} D.~A.,
  {Grillmair} C.~J.,  2001, \aj, 121, 2974

\bibitem[\protect\citeauthoryear{{Lauer}, {Faber}, {Gebhardt}, {Richstone},
  {Tremaine}, {Ajhar}, {Aller}, {Bender}, {Dressler}, {Filippenko}, {Green},
  {Grillmair}, {Ho}, {Kormendy}, {Magorrian}, {Pinkney} \& {Siopis}}{{Lauer}
  et~al.}{2005}]{Lauer05}
{Lauer} T.~R.,   {Faber} S.~M., et al. 2005, \aj, 129, 2138

\bibitem[\protect\citeauthoryear{{Lauer}, {Gebhardt}, {Faber}, {Richstone},
  {Tremaine}, {Kormendy}, {Aller}, {Bender}, {Dressler}, {Filippenko}, {Green}
  \& {Ho}}{{Lauer} et~al.}{2007}]{Lauer07}
{Lauer} T.~R.,   {Gebhardt} K., et al. 2007, \apj, 664, 226

\bibitem[\protect\citeauthoryear{{Lee}, {Park}, {Hwang}, {Arimoto}, {Tamura} \&
  {Onodera}}{{Lee} et~al.}{2010}]{Lee10}
{Lee} M.~G.,   {Park} H.~S., et al. 2010, \apj, 709, 1083

\bibitem[\protect\citeauthoryear{{L{\'o}pez-Sanjuan}, {Balcells},
  {P{\'e}rez-Gonz{\'a}lez}, {Barro}, {Garc{\'{\i}}a-Dab{\'o}}, {Gallego} \&
  {Zamorano}}{{L{\'o}pez-Sanjuan} et~al.}{2010}]{LopezSanjuan10}
{L{\'o}pez-Sanjuan} C.,   {Balcells} M., et al. 2010, \apj, 710, 1170

\bibitem[\protect\citeauthoryear{{Mieske}, {Hilker} \& {Infante}}{{Mieske}
  et~al.}{2004}]{Mieske04}
{Mieske} S.,  {Hilker} M.,    {Infante} L.,  2004, \aap, 418, 445

\bibitem[\protect\citeauthoryear{{Minniti}}{{Minniti}}{1996}]{Minniti96}
{Minniti} D.,  1996, \apj, 459, 175

\bibitem[\protect\citeauthoryear{{Miyazaki}, {Komiyama}, {Sekiguchi},
  {Okamura}, {Doi}, {Furusawa}, {Hamabe}, {Imi}, {Kimura}, {Nakata}, {Okada},
  {Ouchi}, {Shimasaku}, {Yagi} \& {Yasuda}}{{Miyazaki}
  et~al.}{2002}]{Miyazaki02}
{Miyazaki} S.,   {Komiyama} Y., et al. 2002, \pasj,   54, 833

\bibitem[\protect\citeauthoryear{{Muratov} \& {Gnedin}}{{Muratov} \&
  {Gnedin}}{2010}]{Muratov10}
{Muratov} A.~L.,  {Gnedin} O.~Y.,  2010, \apj, 718, 1266

\bibitem[\protect\citeauthoryear{{Napolitano}, {Romanowsky}, {Coccato},
  {Capaccioli}, {Douglas}, {Noordermeer}, {Gerhard}, {Arnaboldi}, {de Lorenzi},
  {Kuijken}, {Merrifield}, {O'Sullivan}, {Cortesi}, {Das} \&
  {Freeman}}{{Napolitano} et~al.}{2009}]{Napolitano09}
{Napolitano} N.~R.,   {Romanowsky} A.~J., et al. 2009, \mnras, 393, 329

\bibitem[\protect\citeauthoryear{{Norris} \& {Kannappan}}{{Norris} \&
  {Kannappan}}{2010}]{Norris10}
{Norris} M.~A.,  {Kannappan} S.~J.,  2010, ArXiv e-prints

\bibitem[\protect\citeauthoryear{{Norris} \& {Kannappan}}{{Norris} \&
  {Kannappan}}{2011}]{Norris11}
{Norris} M.~A.,  {Kannappan} S.~J.,  2011, ArXiv e-prints

\bibitem[\protect\citeauthoryear{{Norris}, {Sharples}, {Bridges}, {Gebhardt},
  {Forbes}, {Proctor}, {Raul Faifer}, {Carlos Forte}, {Beasley}, {Zepf} \&
  {Hanes}}{{Norris} et~al.}{2008}]{Norris08}
{Norris} M.~A.,   {Sharples} R.~M., et al. 2008, \mnras, 385, 40

\bibitem[\protect\citeauthoryear{{O'Sullivan} \& {Ponman}}{{O'Sullivan} \&
  {Ponman}}{2004}]{OSullivan04}
{O'Sullivan} E.,  {Ponman} T.~J.,  2004, \mnras, 349, 535

\bibitem[\protect\citeauthoryear{{Ouchi}, {Shimasaku}, {Okamura}, {Furusawa},
  {Kashikawa}, {Ota}, {Doi}, {Hamabe}, {Kimura}, {Komiyama}, {Miyazaki},
  {Miyazaki}, {Nakata}, {Sekiguchi}, {Yagi} \& {Yasuda}}{{Ouchi}
  et~al.}{2004}]{Ouchi04}
{Ouchi} M.,   {Shimasaku} K., et al. 2004, \apj, 611, 660

\bibitem[\protect\citeauthoryear{{Paturel}, {Petit}, {Prugniel}, {Theureau},
  {Rousseau}, {Brouty}, {Dubois} \& {Cambr{\'e}sy}}{{Paturel}
  et~al.}{2003}]{Paturel03}
{Paturel} G.,   {Petit} C., et al. 2003, \aap, 412, 45

\bibitem[\protect\citeauthoryear{{Peng}, {Ferguson}, {Goudfrooij}, {Hammer},
  {Lucey}, {Marzke}, {Puzia}, {Carter}, {Balcells}, {Bridges}, {Chiboucas},
  {del Burgo}, {Graham} \& et al.}{{Peng} et~al.}{2011}]{Peng11}
{Peng} E.~W.,   {Ferguson} H.~C., et al. 2011, \apj, 730, 23

\bibitem[\protect\citeauthoryear{{Peng}, {Jord{\'a}n}, {C{\^o}t{\'e}},
  {Blakeslee}, {Ferrarese}, {Mei}, {West}, {Merritt}, {Milosavljevi{\'c}} \&
  {Tonry}}{{Peng} et~al.}{2006}]{Peng06}
{Peng} E.~W.,   {Jord{\'a}n} A., et al. 2006, \apj, 639, 95

\bibitem[\protect\citeauthoryear{{Peng}, {Jord{\'a}n}, {C{\^o}t{\'e}},
  {Takamiya}, {West}, {Blakeslee}, {Chen}, {Ferrarese}, {Mei}, {Tonry} \&
  {West}}{{Peng} et~al.}{2008}]{Peng08}
{Peng} E.~W.,   {Jord{\'a}n} A., et al. 2008, \apj, 681, 197

\bibitem[\protect\citeauthoryear{{Pipino}, {D'Ercole}, {Chiappini} \&
  {Matteucci}}{{Pipino} et~al.}{2010}]{Pipino10}
{Pipino} A.,  {D'Ercole} A.,  {Chiappini} C.,    {Matteucci} F.,  2010, \mnras,
  407, 1347

\bibitem[\protect\citeauthoryear{{Proctor}, {Forbes}, {Brodie} \&
  {Strader}}{{Proctor} et~al.}{2008}]{Proctor08}
{Proctor} R.~N.,  {Forbes} D.~A.,  {Brodie} J.~P.,    {Strader} J.,  2008,
  \mnras, 385, 1709

\bibitem[\protect\citeauthoryear{{Proctor}, {Forbes}, {Romanowsky}, {Brodie},
  {Strader}, {Spolaor}, {Mendel} \& {Spitler}}{{Proctor}
  et~al.}{2009}]{Proctor09}
{Proctor} R.~N.,   {Forbes} D.~A., et al. 2009, \mnras,   398, 91

\bibitem[\protect\citeauthoryear{{Proctor} \& {Sansom}}{{Proctor} \&
  {Sansom}}{2002}]{Proctor02}
{Proctor} R.~N.,  {Sansom} A.~E.,  2002, \mnras, 333, 517

\bibitem[\protect\citeauthoryear{{Puzia}, {Kissler-Patig}, {Thomas},
  {Maraston}, {Saglia}, {Bender}, {Goudfrooij} \& {Hempel}}{{Puzia}
  et~al.}{2005}]{Puzia05}
{Puzia} T.~H.,   {Kissler-Patig} M., et al. 2005, \aap, 439, 997

\bibitem[\protect\citeauthoryear{{Reda}, {Proctor}, {Forbes}, {Hau} \&
  {Larsen}}{{Reda} et~al.}{2007}]{Reda07}
{Reda} F.~M.,  {Proctor} R.~N.,  {Forbes} D.~A.,  {Hau} G.~K.~T.,    {Larsen}
  S.~S.,  2007, \mnras, 377, 1772

\bibitem[\protect\citeauthoryear{{Rodionov} \& {Athanassoula}}{{Rodionov} \&
  {Athanassoula}}{2011}]{Rodionov11}
{Rodionov} S.~A.,  {Athanassoula} E.,  2011, \mnras, 410, 111

\bibitem[\protect\citeauthoryear{{Romanowsky}, {Douglas}, {Arnaboldi},
  {Kuijken}, {Merrifield}, {Napolitano}, {Capaccioli} \&
  {Freeman}}{{Romanowsky} et~al.}{2003}]{Romanowsky03}
{Romanowsky} A.~J.,   {Douglas} N.~G., et al. 2003, Science, 301, 1696

\bibitem[\protect\citeauthoryear{{Romanowsky}, {Strader}, {Spitler}, {Johnson},
  {Brodie}, {Forbes} \& {Ponman}}{{Romanowsky} et~al.}{2009}]{Romanowsky09}
{Romanowsky} A.~J.,   {Strader} J., et al. 2009, \aj, 137, 4956

\bibitem[\protect\citeauthoryear{{Sawilowsky}}{{Sawilowsky}}{2007}]{Sawiloswsk%
y07}
{Sawilowsky} S.~S.,  2007, Real Data Analysis.
IAP-Information Age Publishing

\bibitem[\protect\citeauthoryear{{Schlegel}, {Finkbeiner} \&
  {Davis}}{{Schlegel} et~al.}{1998}]{Schlegel98}
{Schlegel} D.~J.,  {Finkbeiner} D.~P.,    {Davis} M.,  1998, \apj, 500, 525

\bibitem[\protect\citeauthoryear{{Schuberth}, {Richtler}, {Hilker}, {Dirsch},
  {Bassino}, {Romanowsky} \& {Infante}}{{Schuberth} et~al.}{2010}]{Schuberth10}
{Schuberth} Y.,   {Richtler} T., et al. 2010, \aap, 513, A52+

\bibitem[\protect\citeauthoryear{{S{\'e}rsic}}{{S{\'e}rsic}}{1963}]{Sersic63}
{S{\'e}rsic} J.~L.,  1963, Boletin de la Asociacion Argentina de Astronomia La
  Plata Argentina, 6, 41

\bibitem[\protect\citeauthoryear{{Sinnott}, {Hou}, {Anderson}, {Harris} \&
  {Woodley}}{{Sinnott} et~al.}{2010}]{Sinnott10}
{Sinnott} B.,  {Hou} A.,  {Anderson} R.,  {Harris} W.~E.,    {Woodley} K.~A.,
  2010, \aj, 140, 2101

\bibitem[\protect\citeauthoryear{{Snyder}, {Hopkins} \& {Hernquist}}{{Snyder}
  et~al.}{2011}]{Snyder11}
{Snyder} G.~F.,  {Hopkins} P.~F.,    {Hernquist} L.,  2011, \apjl, 728, L24+

\bibitem[\protect\citeauthoryear{{Spitler} \& {Forbes}}{{Spitler} \&
  {Forbes}}{2009}]{Spitler09}
{Spitler} L.~R.,  {Forbes} D.~A.,  2009, \mnras, 392, L1

\bibitem[\protect\citeauthoryear{{Spitler}, {Forbes}, {Strader}, {Brodie} \&
  {Gallagher}}{{Spitler} et~al.}{2008}]{Spitler08b}
{Spitler} L.~R.,  {Forbes} D.~A.,  {Strader} J.,  {Brodie} J.~P.,
  {Gallagher} J.~S.,  2008, \mnras, 385, 361

\bibitem[\protect\citeauthoryear{{Spitler}, {Larsen}, {Strader}, {Brodie},
  {Forbes} \& {Beasley}}{{Spitler} et~al.}{2006}]{Spitler06}
{Spitler} L.~R.,   {Larsen} S.~S., et al. 2006, \aj, 132, 1593

\bibitem[\protect\citeauthoryear{{Strader}, {Brodie}, {Cenarro}, {Beasley} \&
  {Forbes}}{{Strader} et~al.}{2005}]{Strader05}
{Strader} J.,  {Brodie} J.~P.,  {Cenarro} A.~J.,  {Beasley} M.~A.,    {Forbes}
  D.~A.,  2005, \aj, 130, 1315

\bibitem[\protect\citeauthoryear{{Taylor}, {Puzia}, {Harris}, {Harris},
  {Kissler-Patig} \& {Hilker}}{{Taylor} et~al.}{2010}]{Taylor10}
{Taylor} M.~A.,   {Puzia} T.~H., et al. 2010, \apj, 712, 1191

\bibitem[\protect\citeauthoryear{{Tonry}, {Dressler}, {Blakeslee}, {Ajhar},
  {Fletcher}, {Luppino}, {Metzger} \& {Moore}}{{Tonry} et~al.}{2001}]{Tonry01}
{Tonry} J.~L.,   {Dressler} A., et al. 2001, \apj,   546, 681

\bibitem[\protect\citeauthoryear{{van den Bergh}}{{van den
  Bergh}}{1982}]{vandenBergh82}
{van den Bergh} S.,  1982, \pasp, 94, 459

\bibitem[\protect\citeauthoryear{{Vazdekis}, {Cenarro}, {Gorgas}, {Cardiel} \&
  {Peletier}}{{Vazdekis} et~al.}{2003}]{V03}
{Vazdekis} A.,  {Cenarro} A.~J.,  {Gorgas} J.,  {Cardiel} N.,    {Peletier}
  R.~F.,  2003, \mnras, 340, 1317

\bibitem[\protect\citeauthoryear{{West}, {C{\^o}t{\'e}}, {Marzke} \&
  {Jord{\'a}n}}{{West} et~al.}{2004}]{West04}
{West} M.~J.,  {C{\^o}t{\'e}} P.,  {Marzke} R.~O.,    {Jord{\'a}n} A.,  2004,
  \nat, 427, 31

\bibitem[\protect\citeauthoryear{{White}}{{White}}{1980}]{White80}
{White} S.~D.~M.,  1980, \mnras, 191, 1P

\bibitem[\protect\citeauthoryear{{Woodley}, {G{\'o}mez}, {Harris}, {Geisler} \&
  {Harris}}{{Woodley} et~al.}{2010}]{Woodley10a}
{Woodley} K.~A.,  {G{\'o}mez} M.,  {Harris} W.~E.,  {Geisler} D.,    {Harris}
  G.~L.~H.,  2010, \aj, 139, 1871

\bibitem[\protect\citeauthoryear{{Woodley}, {Harris}, {Puzia}, {G{\'o}mez},
  {Harris} \& {Geisler}}{{Woodley} et~al.}{2010}]{Woodley10b}
{Woodley} K.~A.,   {Harris} W.~E., et al. 2010, \apj, 708, 1335

\bibitem[\protect\citeauthoryear{{Worthey}}{{Worthey}}{1994}]{Worthey94a}
{Worthey} G.,  1994, \apjs, 95, 107

\bibitem[\protect\citeauthoryear{{Xu}, {Narayanan} \& {Walker}}{{Xu}
  et~al.}{2010}]{Xu10}
{Xu} X.,  {Narayanan} D.,    {Walker} C.,  2010, \apjl, 721, L112

\bibitem[\protect\citeauthoryear{{Yagi}, {Kashikawa}, {Sekiguchi}, {Doi},
  {Yasuda}, {Shimasaku} \& {Okamura}}{{Yagi} et~al.}{2002}]{Yagi02}
{Yagi} M.,   {Kashikawa} N., et al. 2002, \aj, 123, 66

\bibitem[\protect\citeauthoryear{{Yoon}, {Yi} \& {Lee}}{{Yoon}
  et~al.}{2006}]{Yoon06}
{Yoon} S.,  {Yi} S.~K.,    {Lee} Y.,  2006, Science, 311, 1129

\bibitem[\protect\citeauthoryear{{Zepf} \& {Ashman}}{{Zepf} \&
  {Ashman}}{1993}]{Zepf93}
{Zepf} S.~E.,  {Ashman} K.~M.,  1993, \mnras, 264, 611

\end{thebibliography}

\begin{appendix}
\section{Data tables}\label{sec:datatables}
This section contains our measurements for the photometry, kinematics and stellar populations of stars and GCs in NGC~4494.

\begin{table*}
\begin{center}
\caption{Individual values for NGC~4494 galaxy light. Columns 1 and 2 give the position of the individual slits in right ascension and declination (J2000), respectively. The observed velocity moments $V_{\rm obs}$, $\sigma$, $h_3$ and $h_4$ appear in columns 3, 4, 5 and 6. Measured values of the CaT index and [Fe/H] from the method of \citetalias{Foster09}, when available, are shown in columns 7 and 8. We quote the maximum metallicity (i.e., $[Fe/H]=+0.2$) when the measured $CaT_{\rm F09}$ is higher than the metallicity range available in the \citetalias{V03} models. The errors on the quoted $[Fe/H]$ values are based on the index errors only and do not include possible systematics due to the adopted calibration. The full version of this table is available through \mnras.}
\begin{tabular}{cccccccc}
\hline
$\alpha$&$\delta$&$V_{{\rm obs}}$&$\sigma$&$h_3$&$h_4$&$CaT_{{\rm F09}}$&[Fe/H]\\
(hh:mm:ss)&(dd:mm:ss)&(km s$^{-1}$)&(km s$^{-1}$)& & &(\AA)&(dex)\\
(1)&(2)&(3)&(4)&(5)&(6)&(7)&(8)\\
\hline
12:31:28.48&+25:44:22.64&1424.6$\pm$ 32.8&193.5$\pm$ 69.6& 0.06$\pm$0.04& 0.02$\pm$0.06&---&---\\
12:31:27.70&+25:45:00.50&1400.0$\pm$ 13.3&131.2$\pm$ 22.6&-0.01$\pm$0.04& 0.00$\pm$0.05&---&---\\
12:31:29.85&+25:45:25.62&1351.1$\pm$  8.0& 24.5$\pm$ 12.9&-0.03$\pm$0.04&-0.05$\pm$0.09&---&---\\
12:31:28.52&+25:45:36.16&1350.6$\pm$  8.6& 61.3$\pm$ 18.6& 0.01$\pm$0.05& 0.12$\pm$0.08&---&---\\
12:31:35.35&+25:45:38.68&1295.1$\pm$ 23.2&129.8$\pm$ 23.9&-0.03$\pm$0.03&-0.02$\pm$0.04&---&---\\
12:31:32.23&+25:45:41.68&1366.3$\pm$  8.6&  5.3$\pm$ 10.5& 0.00$\pm$0.02&-0.01$\pm$0.04&---&---\\
...&...&...&...&...&...&...&...\\
\hline
\end{tabular}\label{table:halo}
\end{center}
\end{table*}

\begin{table*}
\begin{center}
\caption{Individual spectroscopically confirmed GCs and UCDs around NGC~4494. Columns 2 and 3 give the position in right ascension and declination (J2000), respectively. Columns 4 to 10 present the photometry. The measured recession velocity is given in column 11, while values of the $CaT_{\rm F10}$ index and inferred [Fe/H], when available, are in columns 12 and 13, respectively. The errors on the quoted $[Fe/H]$ are based on the index errors only and do not include systematics e.g. adopted calibration. The full version of this table is available through \mnras.}
\begin{tabular}{ccccccccccccc}
\hline
id&$\alpha$&$\delta$&$g_0$&$r_0$&$i_0$&$(g-i)_0$&$V_0$&$I_0$&$(V-I)_0$&$V_{{\rm obs}}$&CaT$_{\mathrm{F10}}$&[Fe/H]\\
&(hh:mm:ss)&(dd:mm:ss)&(mag)&(mag)&(mag)&(mag)&(mag)&(mag)&(mag)&(km s$^{-1}$)&(\AA)&(dex)\\
(1)&(2)&(3)&(4)&(5)&(6)&(7)&(8)&(9)&(10)&(11)&(12)&(13)\\
\hline
&&&&&&UCD&&&&&&\\
\hline
UCD1&12:31:25.515&+25:46:19.77& 20.17& 19.45& 19.11& 1.06&---&---&---&1281$\pm$ 5&7.64$\pm$0.03&-0.30$\pm$0.01\\
UCD2&12:31:24.643&+25:48:16.09& 20.18& 19.48& 19.07& 1.11&---&---&---&1341$\pm$ 5&7.45$\pm$0.01&-0.38$\pm$0.01\\
UCD3&12:31:27.191&+25:46:01.41& 20.32& 19.66& 19.28& 1.04&---&---&---&1152$\pm$ 5&7.59$\pm$0.01&-0.32$\pm$0.01\\
\hline
&&&&&&GC&&&&&&\\
\hline
GC1&12:31:17.633&+25:47:38.50& 22.73& 22.16& 21.96& 0.77&---&---&---&1215$\pm$ 5&6.21$\pm$0.50&-0.92$\pm$0.22\\
GC1&12:31:17.633&+25:47:38.50& 22.73& 22.16& 21.96& 0.77&---&---&---&1198$\pm$15&---&---\\
GC2&12:31:23.857&+25:45:55.41& 21.25& 20.56& 20.23& 1.01&---&---&---&1502$\pm$ 5&6.69$\pm$0.11&-0.71$\pm$0.05\\
...&...&...&...&...&...&...&...&...&...&...&...&...\\
\hline
\end{tabular}\label{table:GC}
\end{center}
\end{table*}	

\section{The CaT as a metallicity indicator}\label{sec:CaTGCs}

In this section, we focus on the reliability of the CaT as a metallicity indicator for unresolved populations (i.e., GCs and galaxies). We use the data presented in this work to shed some light on two open issues.

First, the \citetalias{V03} SSP models predict that the CaT saturates at high metallicity (i.e., $[Fe/H]=-0.5$). This prediction is consistent with the distribution of the GC data for NGC~1407 in \citetalias{Foster10} but is less clear from Fig. \ref{fig:giCaT} where the relationship between colour and $CaT_{\rm F10}$ appears linear at all probed metallicities. This is consistent with what has been observed in the Galactic GCs \citep{AZ88}. On the other hand, in Fig. \ref{fig:gradient} the measured $CaT_{\rm F09}$ for galaxy spectra scatters about a constant value, which coincides with the saturation value predicted by the \citetalias{V03} models. This may be interpreted as proof of the saturation prediction. However, the galaxy colour gradient (see Fig. \ref{fig:lum_profile}) is shallow at similar radii suggesting that the expected change in $CaT_{\rm F09}$ may be small. In other words, the absence of a CaT radial gradient in the galaxy found in this work and in \citetalias{Foster09} may not prove the \citetalias{V03} saturation prediction. The question remains open.

Second, \citetalias{Foster10} found that the brightest blue and red GCs in NGC~1407 have the same CaT index value despite their wide separation in colour. This cast serious doubt on the reliability of the CaT as a metallicity indicator. Moreover, several fitted GC spectra across the range of GC luminosities probed around NGC~1407 showed Paschen lines. We find no such Paschen lines in the GC spectra for NGC~4494, which cover a comparable range in absolute luminosities. The presence of Paschen features in the GCs around NGC~1407 cannot be confirmed directly on the raw spectra. 

Several possible explanations for these behaviours are put forward in \citetalias{Foster10}, including the presence of hot blue stars such as blue horizontal branch, blue straggler or young stars mainly in the blue GCs whose Paschen lines might be affecting the CaT features. Another possibility is that the CaT may saturate at lower metallicity (i.e. $[Fe/H]\sim-0.8$ dex) than predicted by \citetalias{V03} or perhaps colour does not trace metallicity linearly \citep[e.g.,][]{Yoon06,Peng06,Blakeslee10}. 

Based on the results for NGC~4494 shown in Fig. \ref{fig:giCaT}, there is no clear evidence that the CaT saturates earlier than predicted by \citetalias{V03} or that the CaT behaves non-linearly with colour. In fact, the relationship between $(g-i)_0$ colours and $CaT_{\rm F10}$ is consistent with being linear, albeit with large observational scatter. This suggests that the strange distribution of $CaT_{\rm F10}$ values in the GCs around NGC~1407 may be best explained by the presence of hot blue stars in a significant number of its GCs afterall.

\section{Numerical simulations of the GC system}\label{sec:Kenji}

In this section, we discuss whether the observed kinematics of blue and red GCs around NGC~4494 can be reproduced reasonably well via major merging between two disk galaxies with pre-existing blue and red GCs. 

\subsection{Model description}

In the following, we ignore the presence of the non-rotating intermediate colour GCs. We run {\it dissipationless} simulations (i.e., no gas dynamics and no new formation of stars and clusters) and compare the physical properties of the GC systems in the merger remnants to search for the model that best reproduces the following observed properties of blue and red GCs:
\begin{enumerate}
\item the maximum rotational velocity ($V_{\rm rot,max}$) is as large as 100 km s$^{-1}$ for blue GCs,
\item the central (maximum) velocity dispersion ($\sigma$) is as large as $150$ km s$^{-1}$ for both subpopulations,
\item$V_{\rm rot,max}/\sigma$ is larger than 0.3 for both metal-poor and -rich GCs, and
\item$V_{\rm rot,max}$ is larger in the blue than in red GCs.
\end{enumerate}
We consider that the latest (iv) is one of the key physical characteristics of the GCs around NGC~4494.

Since the numerical methods and techniques we employ for modelling the dynamical evolution of dissipationless mergers between two disks with GCs have been detailed elsewhere \citep{Bekki05,Bekki10}, we give only a brief review focusing on the main particularities of this project here. The progenitor disk galaxies taking part in the merger are given a dark matter halo, bulge, thin exponential disk, as well as GCs initially located in the bulge (hereafter BGCs) and in the halo (HGCs). The total mass and size of the exponential disk (bulge) are $M_{\rm d}$ ($M_{\rm b}$)  and $R_{\rm d}$ ($R_{\rm b}$), respectively. We show results from models where the progenitor disk galaxies are similar to the Galaxy (i.e., $M_{\rm b}=0.167M_{\rm d}$ and $R_{\rm b}=0.2R_{\rm d}$). In this work, the total (virial) mass of the dark matter halo ($M_{\rm dm}$) in a disk galaxy is set to be $9M_{\rm d}$.

The initial distribution of HGCs in the progenitor disk galaxies follows a power-law profile with index $\alpha=-3.5$. The extent of the HGC system ($R_{\rm HGC}$) is assumed to be $3R_{\rm d}$. In other words, the power-law distribution is truncated at $R=3R_{\rm d}$. The half-number radius of the HGC system is set to be $0.29R_{\rm d}$. The initial distribution of BGCs also follows the same power-law but has a truncation radius of $r_{\rm gc} R_{\rm HGC}$, where $r_{\rm gc}$ controls the compactness of the spatial distribution of BGCs. We find that $r_{\rm gc}$ is important for determining the kinematical differences between HGCs and BGCs in the merger remnants. 

In the present models, the inner metal-rich BGCs and outer metal-poor HGCs ultimately become blue and red GCs in NGC 4494, respectively. This is a simplification since additional metal-rich GCs may form from gas dynamics during galaxy merging \citep[e.g.,][]{Bekki02}. Even so, the models may enable us to understand how large amounts of rotation are present in metal-poor and possibly -rich GCs around NGC~4494. For convenience, $V_{\rm rot,max}$ ($\sigma$) of HGCs (blue GCs) and BGCs (red GCs) are referred to as $V_{\rm rot,max,blue}$ (${\sigma}_{\rm blue}$) and $V_{\rm rot,max,red}$ (${\sigma}_{\rm red}$), respectively. Here $R_{\rm d}$ and $M_{\rm d}$ are set to be 13.4 kpc and $4.3 \times 10^{10} {\rm M}_{\odot}$, respectively, to reproduce the observed kinematics of the stars in NGC~4494.

\begin{table*}
\begin{center}
\begin{tabular}{cccccccc}
\hline
Model &$r_{\rm p}$&$e_{\rm p}$&orbit&$V_{\rm rot,max,blue}$&$V_{\rm
rot,max,red}$ &$V_{\rm rot,max,blue}/\sigma_{\rm blue}$&$V_{\rm
rot,max,red}/\sigma_{\rm red}$\\
&($\times R_{\rm d}$)&&&(km s$^{-1}$)&(km s$^{-1}$)&&\\
(1)&(2)&(3)&(4)&(5)&(6)&(7)&(8)\\
\hline
1 & 2.0  & 0.72  & RR & 122 & 72 & 0.78 & 0.54  \\
2 & 1.0  & 1.0  & PP & 47 & 75 & 0.35 & 0.66  \\
3 & 2.0  & 1.0  & PP & 66 & 79 & 0.38 & 0.54  \\
4 & 2.0  & 0.72  & PP & 81 & 111 & 0.57 & 0.86  \\
5 & 2.0  & 0.72  & PR & 73 & 85 & 0.46 & 0.79  \\
\hline
\end{tabular}
\caption{Model (column 1) parameter values and brief summary of results. Column 2 lists the pericenter distance of the merger in units of the disk size $R_{\rm d}$. The orbital eccentricity of the merger is shown in column 3. Description of the orbits in column 4 are coded as follows: ``PP'', ``PR'', and ``RR''  for prograde-prograde, prograde-retrograde, and retrograde-retrograde merging, respectively. Column 5 and 6 shows the maximum rotational velocity of metal-poor and -rich GCs around the remnant, respectively. Columns 7 and 8 list the ratio of the maximum rotational velocity to the central velocity dispersion for metal-poor and -rich GCs, respectively.}\label{table:models}
\end{center}
\end{table*}

\begin{figure}
\includegraphics[width=84mm]{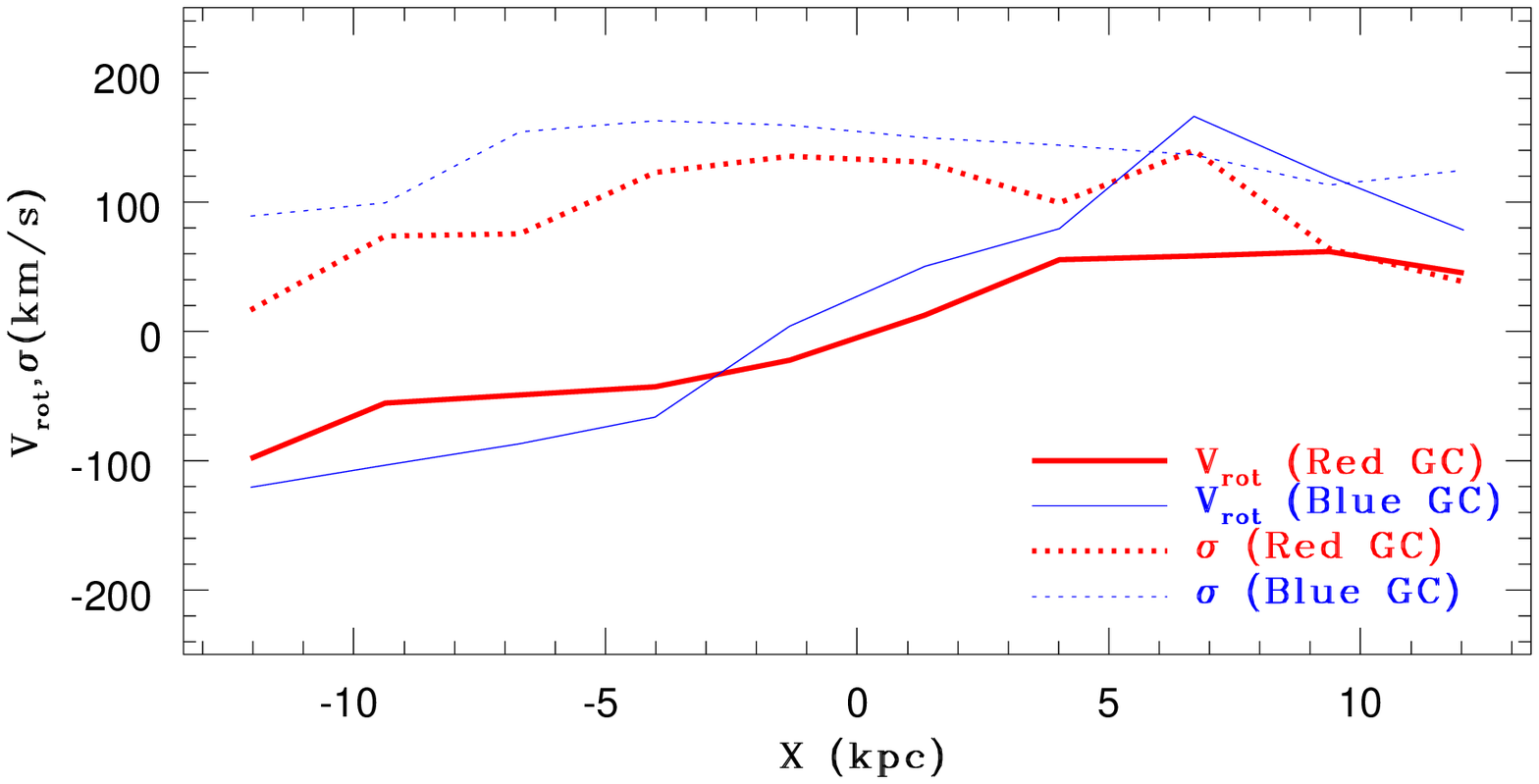}
\caption{Line-of-sight rotational velocities ($V_{rot}$, solid lines) and velocity dispersion ($\sigma$, dotted lines) for blue and red GCs along the $x$-axis in the merger remnant for the fiducial model (Model
1). Here the $y$-components of GC velocities are used for deriving the line-of-sight $V_{rot}$ and $\sigma$ at each bin along the $x$-axis. This figure is available in colour in the online version.}
\label{fig:C1}
\end{figure}

\begin{figure}
\includegraphics[width=84mm]{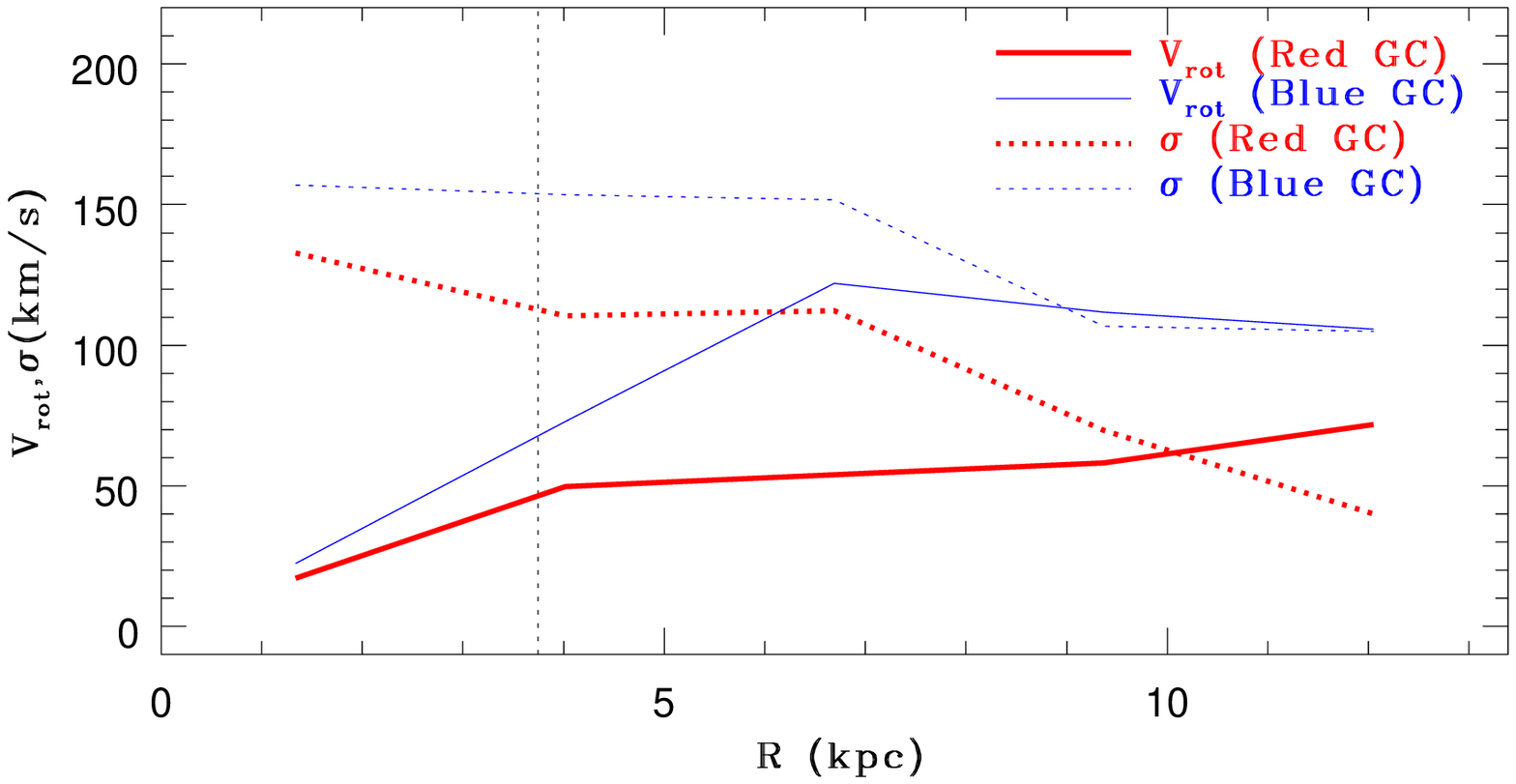}
\caption{Same as Fig. \ref{fig:C1} for the radial dependence directly comparable with the observational results in this work. Vertical dotted line indicates the effective radius of NGC~4494. This figure is available in colour in the online version.}
\label{fig:C2}
\end{figure}

The mass ratio of the two progenitor disks in the simulated mergers is 1 (i.e., equal-mass major merger) and the orbit of the two disks is initially set to be in the $xy$-plane for all models. The pericenter distance ($r_{\rm p}$) and the eccentricity ($e_{\rm p}$) of the merger are free parameters that influence the orbital energy and angular momentum. The spin of each progenitor galaxy is specified by two angles $\theta_{i}$ and $\phi_{i}$. $\theta_{i}$ is the angle between the $z$-axis and the angular momentum vector of the $i^{\rm th}$ progenitor disk. $\phi_{i}$ is the azimuthal angle measured from the $x$-axis to the projection of the angular momentum vector of the $i^{\rm th}$ progenitor disk onto the $xy$-plane.

We mainly show the results for our ``fiducial model'', which has $r_{\rm p}=2R_{\rm d}$, $e_{\rm p}=0.72$, $\theta_{1}=210^{\circ}$, $\theta_{2}=225^{\circ}$, $\phi_{1}=45^{\circ}$, and $\phi_{2}=120^{\circ}$, as it best reproduces the observed four key properties of the NGC~4494 GCs listed above. The fiducial model has an orbital configuration similar to ``retrograde-retrograde'' merging (i.e., the orbital spin axis of the merger is anti-parallel to the intrinsic spin axes of the two disks). We also show the results of the models with ``prograde-prograde'' (``prograde-retrograde'') orbital configurations in which $\theta_{1}=30^{\circ}$ and $\theta_{2}=45^{\circ}$ ($\theta_{1}=30^{\circ}$ and $\theta_{2}=225^{\circ}$) and all other parameters unchanged. Model parameters and some salient results are given in the Table \ref{table:models}.

When estimating the kinematics of the GC system and binned major-axis profiles, the merger remnant is assumed to be viewed near to edge-on. In order to have enough objects in each bin, GCs at corresponding minor axis distances are included. 

\subsection{Results}

Fig. \ref{fig:C1} shows the rotational velocity ($V_{\rm rot}$) and velocity dispersion ($\sigma$) along the $x$-axis of the merger remnant for the fiducial model. The $y$-component of GC velocities for each of 10 bins are used for deriving line-of-sight profiles. Fig. \ref{fig:C1} reveals that both metal-poor and -rich GCs exhibit global rotation ($-12$ kpc $ \le x \le 12$ kpc) albeit with stochastic scatter due to the small number of GCs in each bin. In fact, the apparently rapid change in $V_{rot}$ around $x=7$ kpc is due largely to small number statistics.

In order to compare the simulated profiles with the observed one, we investigate radial ($R$) profiles of the rotational velocity and velocity dispersion (Fig. \ref{fig:C2}). In the fiducial model, the larger amount of orbital angular momentum ($e_{\rm p}=0.72$ and $r_{\rm p}=2R_{\rm d}$) allows the outer components (dark matter halo and HGCs) to acquire a large amount of rotation due to the efficient conversion of orbital angular momentum into spin angular momentum. For $R>R_{\rm e}$ (=3.75kpc), the simulated $V_{\rm rot,blue}$ can be larger than 50 km s$^{-1}$ (see Fig. \ref{fig:C2}) in agreement with the observations presented herein.

As shown in Table \ref{table:models}, the models with parabolic encounters (i.e., models 2 and 3) that initially have a significantly smaller amount of orbital angular momentum do not show large $V_{\rm rot,max,blue}$ compared to the fiducial model. This implies that the observed large rotational velocity ($V_{\rm rot}/\sigma \sim 0.7$) may help constrain the orbits of the progenitors of NGC~4494. Moreover, the small $e_{\rm p}$ in the fiducial model suggests that the progenitor galaxies were either a binary pair of galaxies or two large galaxies dominating a small bound group.

Fig. \ref{fig:C2} also shows that even though the red GCs (BGCs) do have rotation, their rotational amplitude is smaller than that of the blue GCs. In the present models, the final kinematics of red GCs may become comparable to those of the stellar components of the merger remnants as a result of the adopted $r_{\rm gc}=0.1$. The fiducial model has a retrograde-retrograde orbital configuration yielding a final stellar remnant with apparently low rotation ($<50$ km s$^{-1}$) depending on the viewing angle. Fig. \ref{fig:C2} shows that $V_{\rm rot,max,red}$ is significantly smaller than $V_{\rm rot,max,blue}$ as observed in NGC~4494. Moreover, as shown in Table \ref{table:models}, models with prograde-prograde and prograde-retrograde orbital configurations do not yield large kinematic differences between blue and red GCs. This suggests that the orbital configuration of disk-disk major mergers may be the key factor in causing the observed kinematic differences between blue and red GCs.

We thus conclude that the observed kinematics of the GC system around NGC~4494 are {\it broadly} consistent with a formation scenario wherein the remnant galaxy formed via major disk-disk merging, involved large amounts of orbital angular momentum and an initial retrograde-retrograde orbital configuration. An interesting prediction of the fiducial model is that the dark matter halo of NGC~4494 also has a significant amount of global rotation. The predicted maximum line-of-sight rotational velocity of the dark matter halo is $\sim 40$ km s$^{-1}$ and $V_{\rm rot}/\sigma \sim 0.25$. However, the simulated rotation profile of the red GCs ($\sim 50$ km s$^{-1}$) is larger than the observed one ($\sim 20$ km s$^{-1}$) for $R\sim 1-2 R_{\rm e}$ (corresponding to 3.8-7.6 kpc). We do not find models that show comparably low remnant red GC rotation whenever the remnant rotation of the blue GCs is large ($V_{\rm rot,max, blue}\sim 100$ km s$^{-1}$). This could mean that physical processes other than the collisionless major merger scenario explored in this study may be necessary to explain all observed stellar and GC kinematics NGC~4494 in a fully self-consistent manner.

\end{appendix}

\end{document}